\documentclass[%
 reprint,
 amsmath,amssymb,
 aps,floatfix,superscriptaddress,
nofootinbib]
{revtex4-2}

\usepackage{graphicx}
\usepackage{dcolumn}
\usepackage{bm}
\usepackage{epstopdf}
 \usepackage[usenames,dvipsnames]{pstricks}
 \usepackage{epsfig}
 \usepackage{pst-grad} 
 \usepackage{pst-plot} 
 \usepackage[space]{grffile} 
 \usepackage{etoolbox} 
 \makeatletter 
 \patchcmd\Gread@eps{\@inputcheck#1 }{\@inputcheck"#1"\relax}{}{}
 \makeatother
\usepackage{booktabs}
\usepackage{boxhandler}
\usepackage{amsmath}
\usepackage{subcaption}
\usepackage{braket}
\usepackage[cmintegrals]{newtxmath}
\usepackage{multirow}
\usepackage{enumitem}

\begin{document}

\title{Performance of surface codes in realistic quantum hardware}

\author{Antonio deMarti iOlius }
\email{ademartio@tecnun.es}
\affiliation{Department of Basic Sciences, Tecnun - University of Navarra, 20018 San Sebastian, Spain.}
\author{Josu Etxezarreta Martinez}
\affiliation{Department of Basic Sciences, Tecnun - University of Navarra, 20018 San Sebastian, Spain.}
\affiliation{This authors contributed equally.}
\author{Patricio Fuentes}
\affiliation{Department of Basic Sciences, Tecnun - University of Navarra, 20018 San Sebastian, Spain.}
\affiliation{This authors contributed equally.}
\author{Pedro M. Crespo}
\affiliation{Department of Basic Sciences, Tecnun - University of Navarra, 20018 San Sebastian, Spain.}
\author{Javier Garcia-Frias}
\affiliation{Department of Electrical and Computer Engineering, University of Delaware, Newark, DE 19716 USA}

\date{\today}

\begin{abstract}
  Surface codes are generally studied based on the assumption that each of the qubits that make up the surface code lattice suffers noise that is independent and identically distributed (i.i.d.). However, real benchmarks of the individual relaxation ($T_1$) and dephasing ($T_2$) times of the constituent qubits of state-of-the-art quantum processors have recently shown that the decoherence effects suffered by each particular qubit actually vary in intensity. In consequence, in this article we introduce the independent non-identically distributed (i.ni.d.) noise model, a decoherence model that accounts for the non-uniform behaviour of the docoherence parameters of qubits. Additionally, we use the i.ni.d model to study how it affects the performance of a specific family of Quantum Error Correction (QEC) codes known as planar codes. For this purpose we employ data from four state-of-the-art superconducting processors: ibmq\_brooklyn, ibm\_washington, Zuchongzhi and Rigetti Aspen-M-1. Our results show that the i.i.d. noise assumption overestimates the performance of surface codes, which can suffer up to $95\%$ performance decrements in terms of the code pseudo-threshold when they are subjected to the i.ni.d. noise model. Furthermore, we consider and describe two methods which enhance the performance of planar codes under i.ni.d. noise. The first method involves a so-called re-weighting process of the conventional minimum weight perfect matching (MWPM) decoder, while the second one exploits the relationship that exists between code performance and qubit arrangement in the surface code lattice. The optimum qubit configuration derived through the combination of the previous two methods can yield planar code pseudo-threshold values that are up to $650\%$ higher than for the traditional MWPM decoder under i.ni.d. noise.

\end{abstract}

\maketitle

\section{Introduction}
Quantum computing heralds the arrival of a new era in computer science where problems that are not within reach for classical computers will become tractable. The principal tenet of quantum computing is to design ingenious algorithms that are capable of exploiting the superposition property of quantum states, which ultimately allows them to consider large portions of problem solution spaces concurrently. Generally, quantum computers are understood as ensembles of qubits, two-level coherent quantum systems that can be employed to leverage the quantum mechanical property of superposition. It must be mentioned, however, that quantum processors can also be constructed using more complex and higher discrete-level coherent quantum systems, known as qudits \cite{qudits,qudits2}, or even continuous intervals \cite{CVQC,CVQC2}. At the time of writing, significant efforts and resources are being destined towards the construction of a large-scale universal fault-tolerant quantum computer. Nonetheless, even though substantial progress has been made in the field in recent years, machines with the capacity to fulfil the complete promise of quantum computing remain, as of yet, nonexistent. 

The main cause for this is that currently existing quantum computers are too noisy to run sophisticated quantum algorithms reliably. The noise of a qubit is generally defined by its decoherence parameters (relaxation time, $T_1$, and dephasing time, $T_2$), which are measures of how long the qubit can maintain its coherence and, thus, be employed reliably to perform calculations \cite{NielsenChuang,josurev,TVQC}. Unfortunately, present-day qubits lack sufficiently long coherence times to enable reliable quantum computing. This occurs because the coherence time of modern qubits is too short relative to the amount of time that is required to interact with them. Qubits are manipulated through the action of quantum gates, whose application consumes much of the coherence time of qubits, and makes it difficult to perform complex and reliable quantum calculations. While the coherence time of qubits varies depending on how they are built (qubits constructed with ion traps present decoherence times in the order of seconds while these times are in the order of hundreds of microseconds for superconducting qubits), so do their gate operation times (superconducting quantum gates are much faster than ion trap quantum gates). For this reason, regardless of which technology is used to implement them, currently existing qubits will suffer from similar noise processes. 

Quantum states experience coherence losses as a result of the unwanted interactions that qubits have with their environment. These interactions arise through myriads of physical mechanisms, many of them unavoidable, and they are all grouped under the same term: decoherence. In fact, other sources of errors in quantum computers, such as faulty gates or inaccurate measurements, can also fall under the umbrella of decoherence. Thus, within the abstraction that decoherence provides to represent quantum noise, the technological odyssey of building a reliable quantum computer can be simply summarized as the search for strategies to effectively fight the effects of decoherence. It is in answer to this challenge that the discipline of QEC arose, to study the phenomenon of decoherence and to design strategies to protect qubits from quantum noise. Similar to what is done in the classical computing framework, QEC strategies, known as QEC codes (QECCs), employ additional qubits to protect quantum information from the impact of decoherence-related effects. In fact, thanks to certain similarities between the classical and quantum computing paradigms, QECCs can be built from existing classical codes. This is achieved by casting existing groups of classical codes into the QEC framework by means of the well-known stabilizer formalism \cite{QSC}. In consequence, many classical-inspired QEC code families like Quantum Low Density Parity Check (QLDPC) codes \cite{bicycle,qldpc15,patrick,patrick2} or Quantum Turbo Codes (QTC) \cite{QTC,EAQTC,josu,josu2}, among many others, are already being studied.

However, because many of these code families require large numbers of fully connected qubits to successfully battle quantum noise,  practical QEC solutions for present day quantum computers are generally based on surface codes, a different type of stabilizer code that is not based on previously existing classical codes \cite{qecsim,toric}. Surface codes are constructed by encoding logical quantum states (those qubits that contain the information that will be processed) into two-dimensional lattices of physical qubits. Particular qubits in the lattice act as measurement qubits that can be used to extract quantum syndromes, binary vectors that enable error diagnosis of qubits while avoiding direct measurement of quantum states, effectively allowing us to perform the appropriate recovery operations without destroying the superposition state of the logical qubits.

Research within the field of QEC, including the surface code niche, typically assumes that the qubits that make up error correction codes will suffer decoherence-related errors that can all be described by the same probability distribution, i.e, that in each error correction round every qubit experiences noise defined by an independent identically distributed process (i.i.d.) \cite{bicycle,qldpc15,patrick,patrick2,QTC,EAQTC,josu,josu2,qecsim,toric}. However, experimental results have shown that the relaxation and dephasing times of qubits in real quantum hardware are actually significantly different \cite{IBMqexp,Wash,Broo,Zuchongzhi,Aspen11}, with drastic variation in the decoherence parameter values of particular qubits. Given that the decoherence-induced noise experienced by superconducting qubits is characterized by their corresponding $T_1$ and $T_2$ times \cite{josurev,TVQC}, the data from real quantum processors suggests that studying surface codes under the i.i.d. qubit noise assumption does not provide the most accurate portrayal of their performance. For this reason, in this article we introduce the independent non-identically distributed (i.ni.d) qubit-noise model as a way to capture and reflect the differences in the decoherence parameters of real qubits. Additionally, we use the aforementioned model to study how the performance of surface codes changes over the proposed i.ni.d model. The primary difference between the i.ni.d docoherence model and the conventional i.i.d. assumption is that the former model considers that each particular qubit of the surface code lattice has its own decoherence defining parameters ($T_1$ and $T_2$) and will experience different noise levels, whereas the latter model (the i.i.d. model) assumes that all qubits are defined by the same decoherence parameters and hence every physical qubit is subjected to the same noise level.

To provide a realistic view of the impact that the more accurate i.ni.d. model can have on the performance of surface codes, we have used the values of the relaxation and dephasing times of modern quantum processors (ibm\_washington \cite{Wash}, ibmq\_brooklyn \cite{Broo}, Zuchongzhi \cite{Zuchongzhi} and Aspen-M-1 \cite{Aspen11}) in our simulations. Our results show that the i.i.d. qubit-noise assumption provides an overly optimistic portrayal of the performance of surface codes when they operate on real hardware. Fortunately, we also show that methods that remarkably enhance performance (at some points surpassing that which would be expected based on i.i.d results) exist.

\section{Preliminaries} \label{sec:prelims}

\subsection{Planar Codes}

Surface codes are a family of quantum error correction codes whose constituent qubits are laid out on a two-dimensional lattice. The qubits within the code interact locally, i.e., only with their nearest neighbours. Depending on the geometry of the lattice, different types of surface codes can be constructed. For our work we consider planar codes arranged over the fundamental lattice in surface code research, the square lattice \cite{toric,surfaceReview}. In this particular configuration, the planar code qubits are arranged on the center, edges and vertices of a square lattice.

There are two different types of qubits within the square lattice that defines a planar code: the data qubits, which are located on the edges of the lattice and encode the quantum state of the code, and the measurement qubits, which are continuously initialized and measured in order to obtain information regarding errors that may have arisen. Measurement qubits interact locally with their nearest data qubit neighbours and will act differently depending on where they are located. Based on how they act on their neighbouring data qubits, we can also classify measurement qubits into two separate groups. On one hand we have vertex qubits or ``measure-X'' qubits, which force their surrounding data qubits into an eigenstate of the operator product $\mathrm{X}_1\mathrm{X}_2\mathrm{X}_3\mathrm{X}_4$, where $\mathrm{X}_i$ is the Pauli $\mathrm{X}$ operator for a specific qubit $i$ and $1$, $2$, $3$ and $4$ are the nearest neighbour data qubits of the considered measurement qubit. On the other hand, we also have plaquette qubits or ``measure-Z'' qubits, which force the surrounding data qubits into an eigenstate of the operator product $\mathrm{Z}_1\mathrm{Z}_2\mathrm{Z}_3\mathrm{Z}_4$, where $\mathrm{Z}$ is the Pauli $\mathrm{Z}$ operator. These concepts are reflected in FIG. \ref{planarcode}, which portrays a graphical representation of a $7\times7$ planar code. Notice that the boundaries are not equal, i.e, the top and bottom lattice boundaries apply vertex measurement qubits while the right and left lattice boundaries apply plaquette measurement qubits. 

The code is initialized by collapsing all measurement qubits so that the data qubits are forced into an eigenstate of all their operator products. The resulting state is known as the quiescent state \cite{surfaceReview}. Once the quiescent state has been obtained, subsequent measurements of measurement qubit states will remain unchanged since the data qubits are in an eigenstate of their operator products \cite{surfaceReview}. For this reason, any change in the measurement of any measure-X or measure-Z qubit will imply that the code is no longer in the quiescent state it was initialized in. Moreover, since the qubits interact locally with their nearest neighbours, this means that one or three of its adjacent data qubits have experienced a Pauli error. Measure-X errors will be susceptible to $\mathrm{Z}$ and $\mathrm{Y}$ errors, while measure-Z errors will be susceptible to $\mathrm{X}$ and $\mathrm{Y}$ errors, as shown in eq. \ref{anticomn}:

\begin{figure}[h]
    \centering
    \includegraphics[width=80mm,scale=1]{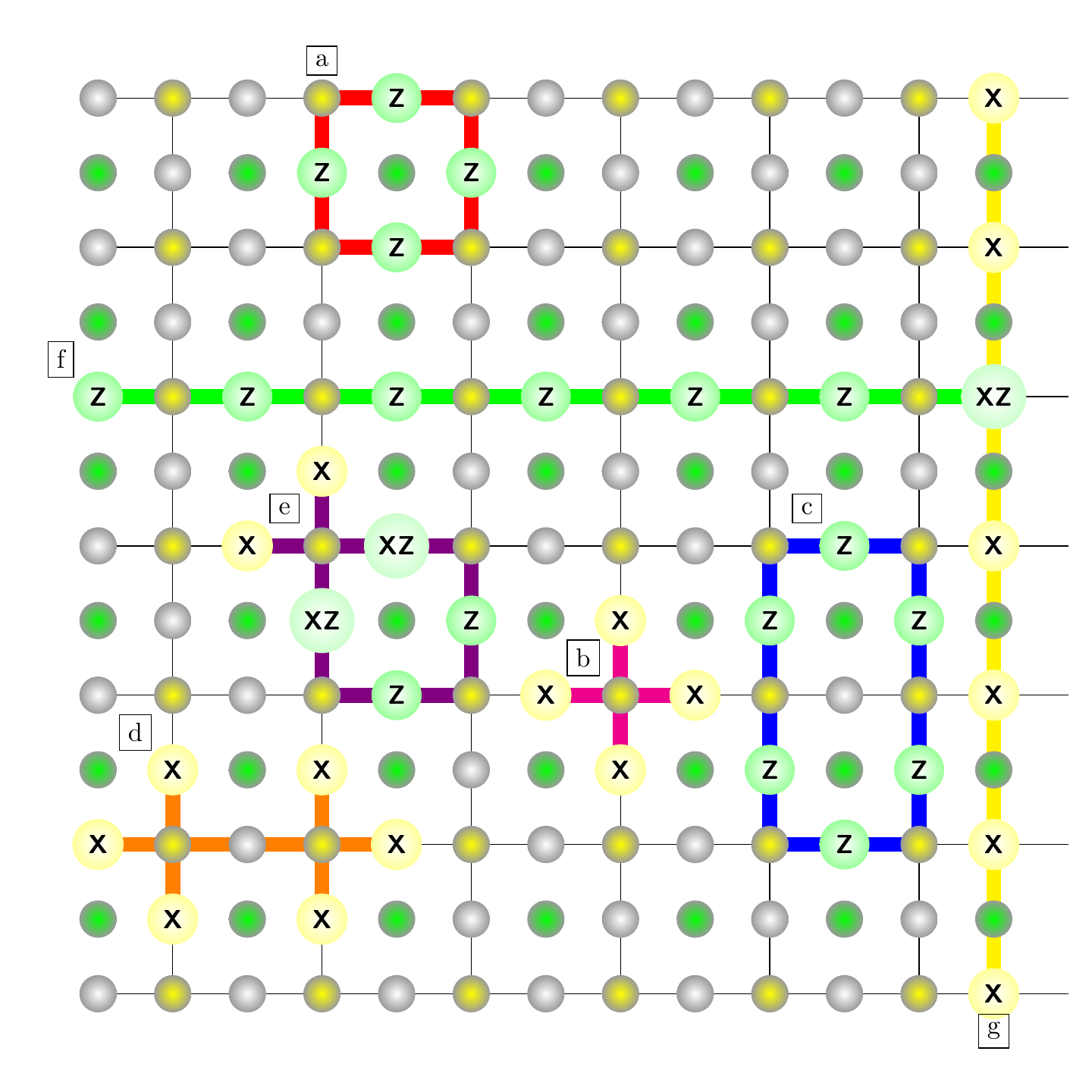}
    \caption{\textbf{Graphical representation of a $7\times 7$ planar code.} The data qubits, the measure-X qubits and the measure-Z qubits are depicted by white, yellow, and green dots, respectively. Data qubits that suffer $\mathrm{X}$, $\mathrm{Z}$ and $\mathrm{XZ}$ operators are portrayed by light green, light yellow and lighter green dots. The action of various stabilizer elements is highlighted using different colour patterns: \textbf{a} action of a measure-Z qubit. \textbf{b} action of a measure-X qubit. \textbf{c} combination of two adjacent plaquette-plaquette stabilizers. \textbf{d} combination of two adjacent vertex-vertex stabilizers. \textbf{e} combination of two adjacent vertex-plaquette stabilizers. \textbf{f} $\mathrm{Z}_L$ operator. \textbf{g} $\mathrm{X}_L$ operator.}
    \label{planarcode}
\end{figure}

\begin{equation}
\label{anticomn}
\begin{aligned}
    \mathrm{X}_a\mathrm{X}_b\mathrm{X}_c\mathrm{X}_d\mathrm{Z}_a \ket{\psi} &= -\mathrm{Z}_a\mathrm{X}_a\mathrm{X}_b\mathrm{X}_c\mathrm{X}_d   \ket{\psi},
    \\
    \mathrm{Z}_a\mathrm{Z}_b\mathrm{Z}_c\mathrm{Z}_d\mathrm{X}_a \ket{\psi} &= - \mathrm{X}_a \mathrm{Z}_a\mathrm{Z}_b\mathrm{Z}_c\mathrm{Z}_d \ket{\psi},
    \\
    \mathrm{X}_a\mathrm{X}_b\mathrm{X}_c\mathrm{X}_d \mathrm{Y}_a\ket{\psi} &= \mathrm{X}_a\mathrm{X}_b\mathrm{X}_c\mathrm{X}_d i\mathrm{X}_a\mathrm{Z}_a \ket{\psi}
    \\
    &= -i\mathrm{X}_a\mathrm{Z}_a\mathrm{X}_a\mathrm{X}_b\mathrm{X}_c\mathrm{X}_d  \ket{\psi}
    \\
    &= -\mathrm{Y}_a\mathrm{X}_a\mathrm{X}_b\mathrm{X}_c\mathrm{X}_d   \ket{\psi},
\end{aligned}
\end{equation}
where $a,b,c,d$ refer to the four surrounding measurement qubits, and $\mathrm{X,Y,Z}$ are the non-identity Pauli matrices. As a result, collapsing the measurement qubits serves to extract the syndrome associated to the error that has taken place at any given instance.

Planar codes like the one depicted in FIG. \ref{planarcode} encode one logical qubit. Logical operators modify the logical state of the surface code in a non-trivial manner while remaining within the codespace (the resulting state commutes with all the stabilizers). We label the Pauli logical operators of the code as $\mathrm{X}_L$, $\mathrm{Y}_L$ and $\mathrm{Z}_L$. The logical operators $\mathrm{X}_L$ and $\mathrm{Z}_L$ can be applied by manipulating the degrees of freedom of the surface code. This is shown in FIG. \ref{planarcode}. Consider the set of $\mathrm{Z}$ operators that traverse the entire planar code lattice horizontally (green line in the figure). These operators commute with all the stabilizer generators of the code, hence they end up forming a $\mathrm{Z}_L$ operator. Similarly, a series of adjacent $\mathrm{X}$ operators that cross the surface code lattice vertically end up forming a $\mathrm{X}_L$ operator. We refer to adjacent $\mathrm{Z}$ operators that traverse the edges of the lattice as chains while adjacent $\mathrm{X}$ operators within the centre of lattice squares are known as co-chains. When the combination of logical operators $\mathrm{X}_L\mathrm{Y}_L$ is applied, $\mathrm{X}$ and $\mathrm{Z}$ operators coincide on the same qubit. We construct $\mathrm{Y}_L$ operators as the product of the aforementioned $\mathrm{X}_L$ and $\mathrm{Z}_L$ logical operators: $\mathrm{Y}_L=\mathrm{Z}_L \mathrm{X}_L$. Also, $\mathrm{X}_L^2=\mathrm{Y}_L^2=\mathrm{Z}_L^2=\mathrm{I}$, since the square of any of these logical operators can be written in terms of the stabilizer generators and so they will not modify the state of the code. Whenever the noise introduced by an $n$-qubit Pauli channel results in the formation of chains or co-chains and the creation of a logical operator, a logical error will take place. This modifies the state of the logical qubit in a non-trivial manner but results in a trivial quantum syndrome when collapsing the measurement qubits (recall that logical operators preserve the codespace). More specifically,  the combination of the operator induced by the quantum noise and the recovery operator can form logical bit-flip errors ($\mathrm{X}_L$), logical phase-flip errors ($\mathrm{Z}_L$) and logical bit-and-phase-flip errors ($\mathrm{Y}_L$). Decoding failures in which wrong error estimates that still result in a non-trivial syndrome are also considered to be logical errors. All in all, logical errors act harmfully on our encoded quantum states and make it difficult to maintain the logical qubit in the desired original quantum state. Avoiding and minimizing the likelihood of chain and co-chain formation is critical for the planar code to successfully correct errors.

\subsection{Probability pseudo-threshold}

\begin{figure}[h]
    \centering
    \includegraphics[width=0.8\columnwidth,scale=1]{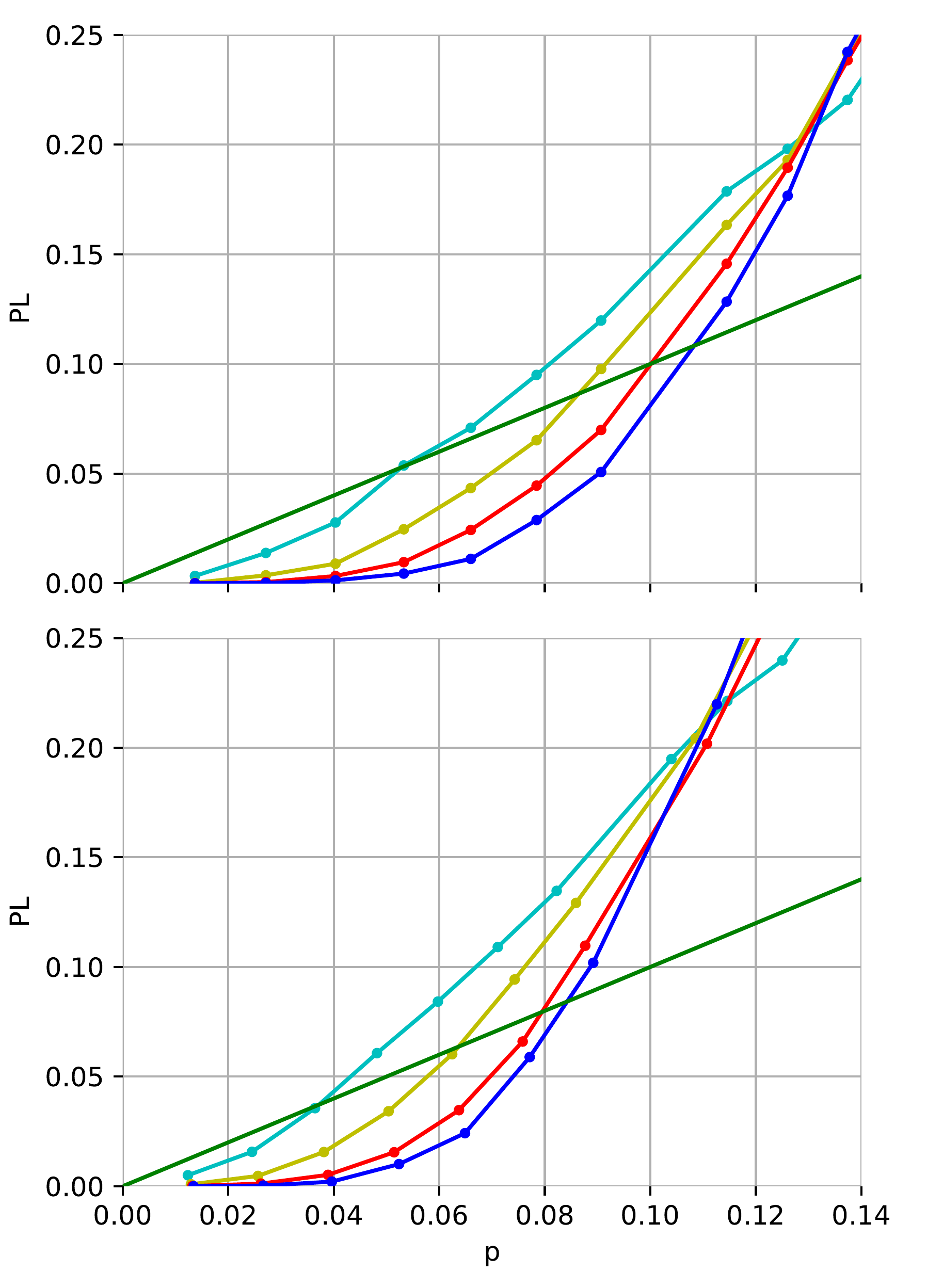}
    \caption{\textbf{Logical error rates of planar codes for the i.i.d. and i.ni.d. models.} \textbf{a} Planar codes operating over the i.i.d. noise model. \textbf{b} Planar codes operating over the i.ni.d. noise model constructed with data from the ibm\_washington hardware \cite{Wash}. The green line represents the performance of the uncoded system, i.e. $P_L(p_{pth})=p_{pth}$. Note that the x-axis represents the probability that a Pauli error occurs $p=p_\mathrm{X} + p_\mathrm{Y} + p_\mathrm{Z}$ and represents the value of $p$ obtained for the mean values of $T_1$ and $T_2$ (for the i.ni.d. channel, since each value of $p$ is different for each qubit, we use the overall mean for the plot).}
    \label{indivvsaver}
\end{figure}

The quantum code probability threshold indicates the maximum physical error probability at which increasing the distance (the distance of a planar code is dictated by its number of physical qubits, i.e, making the code larger increases its distance) of the code lowers the logical error probability \cite{NielsenChuang}. Thus, if the physical error probability is below this threshold, $p<p_{th}$, adding qubits to the error correction code will result in better code performance. Determining the code threshold is one of the primary ways to benchmark the performance of surface codes in the literature when i.i.d. noise is considered \cite{surfaceReview}. However, as can be seen in FIG.\ref{indivvsaver}, when the differences in the $T_1$ and $T_2$ values of physical qubits that make up quantum systems are accounted for, there is no longer a threshold physical probability value at which the performance of surface codes with different distances crosses. While in the top subfigure of FIG.\ref{indivvsaver} all the distance curves converge onto the same point, we can see in the bottom subfigure that this convergence region has spread out \footnote{Notice that different distance curves cross at different $p$ values.}.

To circumvent this issue and maintain our ability to benchmark the quality of planar codes over the i.ni.d channel, we apply the so-called probability pseudo-threshold \cite{pseudoth,pseudoth2}. The code pseudo-threshold is the physical error probability at which the logical error rate meets the physical error probability, $P_L(p_{pth})=p_{pth}$. More specifically, it is the physical error rate at which a code of distance $d$ performs as well (or as poorly) as an uncoded system. This physical error probability value can be thought of as the point beyond which building the error correction code will defy its own purpose, as the code will fail more frequently than the system it seeks to correct. The $p_{pth}$ of FIG.\ref{indivvsaver} can be seen in the intersections of each distance curve with the green line.

\subsection{MWPM Decoder}

 In surface codes, decoding a syndrome is equivalent to finding paths between the stabilizer generators whose syndrome elements have been triggered. We employ the Minimum Weight Perfect Matching (MWPM) decoder to estimate the errors that have taken place based on the measured syndrome \cite{surfaceReview,MWPM,MWPMFowler}. Multiple paths are associated to a given syndrome and the task of the decoder is to produce an estimate of the path that associated to the error that has the highest probability of taking place. The MWPM decoder finds a solution to this problem by searching for the minimum-weight path within the lattice. In graph theory, the MWPM problem is described as that of finding a matching (a set of edges without common vertices) whose weight sum (the sum of edge weights) is minimized. The term perfect refers to the fact that the matching matches all vertices of the graph \cite{MWPM}. The lattice of a surface code that has suffered an error can be converted to a complete graph, where the generators with non-trivial syndrome components are the nodes \cite{MWPMFowler}. The edges between the vertices have a weight equal to the minimum number of qubits between them. In this manner, by deriving a graph associated to the code lattice, solving the MWPM problem by finding the path of minimum weight over this graph serves to produce an estimate of the most likely error that the code has suffered \cite{MWPM}.

The MWPM algorithm is an effective method for low physical error probabilities, but numerical methods have shown that at a given threshold of around $p* = 10.3\%$ its performance drops severely. This occurs because at such noise levels, the error operator with the highest probability (the decoder estimate) does not usually belong to the error equivalence class with the highest probability \cite{degenMWPM}. This is conventionally known as the degeneracy of quantum error correction codes \cite{degen}.

\section{The independent non-identically distributed decoherence model}

The decoherence induced errors in superconducting qubits arise mainly from the combination of energy relaxation and pure dephasing processes. The so-called combined amplitude and phase damping channel, $\mathcal{N}_\mathrm{APD}$, provides a fairly complete mathematical abstraction of the aforementioned processes that corrupt quantum information \cite{josurev,TVQC}. Simulating the combined amplitude and phase damping channel requires an exponential amount of resources, and so it is impractical to track the effects of this channel through classical means. However, by invoking  the well-known twirling technique, we can obtain a more symmetric version of the amplitude and phase damping channel that preserves the noise dynamics of the original quantum channel and that can also be simulated on a classical machine \cite{josurev}. Additionally, it has been shown that correctable codes constructed for this twirled version of the channel will also be correctable codes for the original channel (up to unitary correction) \cite{twirlcorrectable}. Thus, it is a common convention in the field of QEC for quantum coding theorists to work with the twirled approximated channels in order to design and simulate QECCs. In particular, in this work we consider the Pauli twirled approximation (PTA) of the $\mathcal{N}_\mathrm{APD}$ channel, denoted by  $\mathcal{N}_\mathrm{APDPTA}$, as our primary decoherence model. This twirled approximation is obtained by averaging the original channel over the set of unitaries defined by the Pauli group, which results in a Pauli channel, $\mathcal{N}_\mathrm{APDPTA}(\rho)=(1-p_\mathrm{X}-p_\mathrm{Y}-p_\mathrm{Z})\rho + p_\mathrm{X} \mathrm{X}\rho\mathrm{X} + p_\mathrm{Y}\mathrm{Y}\rho\mathrm{Y}+p_\mathrm{Z}\mathrm{Z}\rho\mathrm{Z}$, with the following probabilities \cite{josurev}
\begin{equation}\label{eq:PTAprobs}
\begin{split}
& p_\mathrm{I} = 1 - p_\mathrm{X} - p_\mathrm{Y} - p_\mathrm{Z}, \\
& p_\mathrm{X} = p_\mathrm{Y} = \frac{1}{4}(1 - \mathrm{e}^{-\frac{t}{T_1}})\text{ and} \\
& p_\mathrm{Z} = \frac{1}{4}(1 + \mathrm{e}^{-\frac{t}{T_1}} - 2\mathrm{e}^{-\frac{t}{T_2}}),
\end{split}
\end{equation}
where $\mathrm{I,X,Y,Z}$ are the identity, bit-flip, bit-and-phase-flip and phase-flip Pauli matrices, respectively. Notice how the probabilities that the Pauli operators have of taking place are directly related to the relaxation time, $T_1$, and the dephasing time, $T_2$. 

The literature on quantum error correction usually considers that each of the qubits of the system is subjected to a noise operation which is independent and identically distributed \cite{bicycle,qldpc15,patrick,patrick2,QTC,EAQTC,josu,josu2,qecsim,toric}. This implies that each particular qubit will have the same probability of suffering a particular Pauli operator within a given error correction block. We refer to this model as i.i.d. noise. Against this backdrop, assuming that we have an $n$-qubit system, the i.i.d. channel that arises can be described as
\begin{equation}\label{eq:iidchan}
\begin{split}
\mathcal{N}^{(n)}_\mathrm{APDPTA}(\rho) &= \mathcal{N}_\mathrm{APDPTA}^{\otimes n}(\rho,\mu_{T_1},\mu_{T_2})\\  &= \sum_{\mathrm{A}\in\{\mathrm{I,X,Y,Z}\}^{\otimes n}} p_\mathrm{A}(\mu_{T_1},\mu_{T_2}) \mathrm{A}\rho\mathrm{A},
\end{split}
\end{equation}
where $\mathrm{A} = \mathrm{A}_{1}\otimes\cdots\otimes\mathrm{A}_{n-1}\otimes\mathrm{A}_n$ with $\mathrm{A}_j\in\{\mathrm{I,X,Y,Z}\}$ denotes each of the possible $n$-fold Pauli error operators, with probability distribution $p_\mathrm{A}(\mu_{T_1},\mu_{T_2})$
\begin{equation}\label{eq:iidprob}
p_{\mathrm{A}}(\mu_{T_1},\mu_{T_2}) = \prod_{j=1}^{n}p_{\mathrm{A}_j}(\mu_{T_1},\mu_{T_2}),
\end{equation}
with $p_{\mathrm{A}_j}(\mu_{T_1},\mu_{T_2})$ described by equation \eqref{eq:PTAprobs}, and where $\mu_{T_1}$ and $\mu_{T_2}$ represent the mean values of the relaxation and dephasing times averaged accross $n$ qubits.

However, real $T_1$ and $T_2$ measurements for various modern experimental superconducting processors disprove the assumption that all the qubits of a superconducting processor are subjected to the same level of noise \cite{IBMqexp,Wash,Broo,Zuchongzhi,Aspen11}. The actual values of these parameters vary substantially from qubit to qubit (these differences can sometimes reach an order of magnitude), which, naturally, cannot be reflected by the noise model of \eqref{eq:iidchan} (See Supplementary Note 1 for more details). For this reason, we must come up with a noise model that can account for such qubit behavioural differences. Therefore, we will consider that the errors experienced by each of the qubits of quantum systems are governed by probability distributions that are independent, and non-identically (i.ni.d.) distributed. This means that the values of $p_\mathrm{X},p_\mathrm{Y},p_\mathrm{Z}$ for each of the qubits within the system will be different. Thus, we will refer to this model as i.ni.d noise. Following this rationale, these i.ni.d. $n$-qubit channels will have the following structure

\begin{equation}\label{eq:inidchan}
\begin{split}
&\mathcal{N}^{(n)}_\mathrm{APDPTA}(\rho) = \bigotimes_{i=1}^n\mathcal{N}_\mathrm{APDPTA}(\rho,\mu_{T_1^j},\mu_{T_2^j})\\  &= \sum_{\mathrm{A}\in\{\mathrm{I,X,Y,Z}\}^{\otimes n}} p_\mathrm{A}(\{T_1^j\}_{j=1}^n,\{T_2^j\}_{j=1}^n) \mathrm{A}\rho\mathrm{A},
\end{split}
\end{equation}
where $\mathrm{A} = \mathrm{A}_{1}\otimes\cdots\otimes\mathrm{A}_{n-1}\otimes\mathrm{A}_n$ with $\mathrm{A}_j\in\{\mathrm{I,X,Y,Z}\}$ denotes each of the possible $n$-fold Pauli error operators with probability distribution $p_\mathrm{A}(\{T_1^j\}_{i=1}^n,\{T_2^j\}_{j=1}^n)$
\begin{equation}\label{eq:inidprob}
p_\mathrm{A}(\{T_1^j\}_{j=1}^n,\{T_2^j\}_{j=1}^n) = \prod_{j=1}^{n}p_{\mathrm{A}_j}(T_1^j,T_2^j),
\end{equation}
with $p_{\mathrm{A}_j}(T_1^j,T_2^j)$ described by equation \eqref{eq:PTAprobs}.

Finally, it is important to state that there are other sources of errors that do not stem from environmental qubit interactions that may also be taken into account to study surface codes. These errors are caused by faulty implementations of gates (gate errors) and measurements that are inaccurate (measurement errors) \cite{surfaceReview}. Considering these additional sources of corruption is important to construct surface codes that are effective, but it is outside the scope of this work. Herein, we limit our analysis to studying the impact that including the differences in qubit $T_1$ and $T_2$ values can have on the performance of error correction codes.

\section{The reweighted MWPM}

Conventional Minimum Weight Perfect Matching decoding suffers harsh performance degradation when it is applied to decode a surface code exposed to i.ni.d. noise (this is shown further on in the Results section). Primarily, this loss stems from the fact that the qubits of the code are no longer identical, which means that some will perform better than others. The standard MWPM decoder considers that the minimum weight set of chains matching the measurement qubits with 1-syndrome contribution are the most probable, where all the edges are of the same weight. Unfortunately, this no longer holds when the physical qubits of the code exhibit different error parameters. Nonetheless, it is possible to substantially minimize the degradation suffered by MWPM decoders over the i.ni.d channel by applying a set of modifications to the decoding algorithm.

\begin{figure*}[htp]
    \centering
    \includegraphics[width=0.6\columnwidth,height = 1.85in]{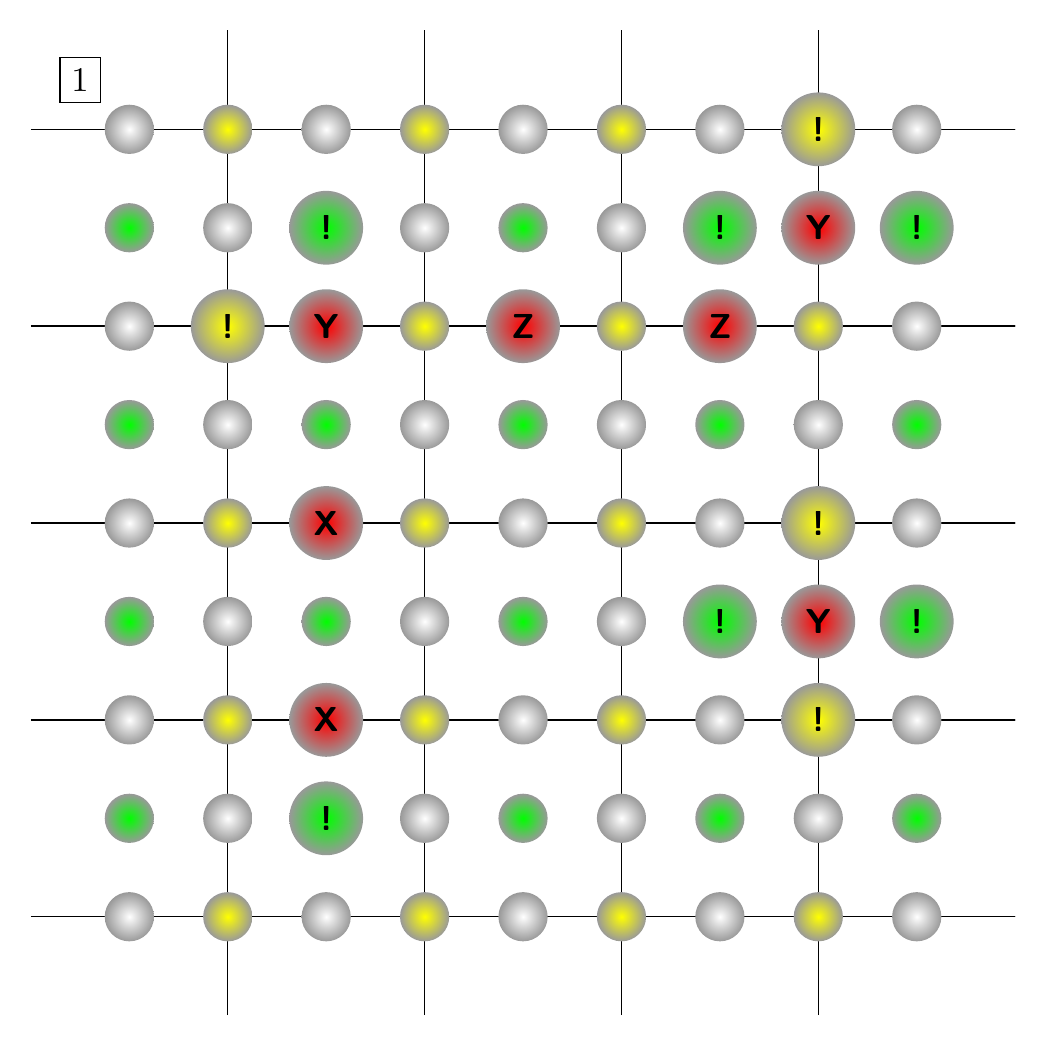}
    \includegraphics[width=0.6\columnwidth,height = 1.85in]{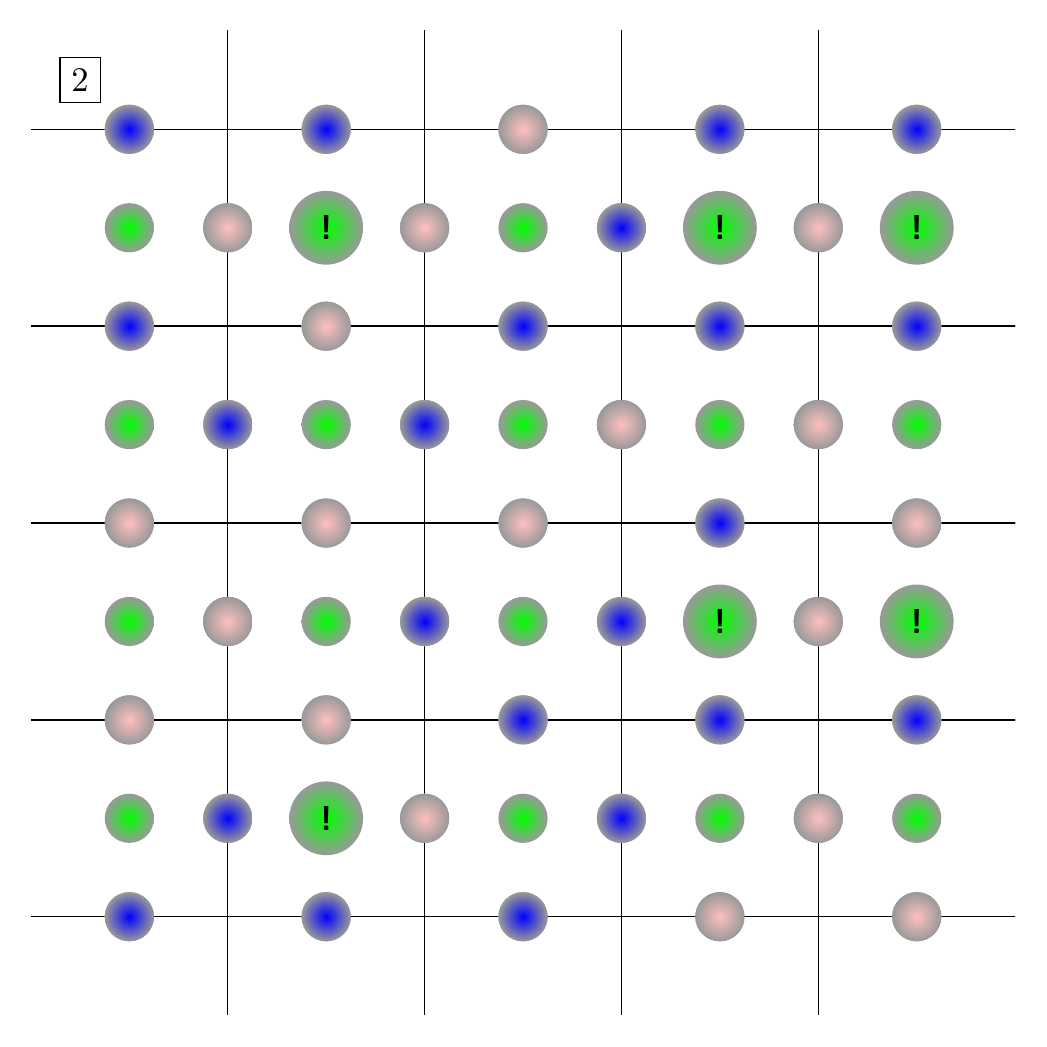}
    \includegraphics[width=0.6\columnwidth,height = 1.85in]{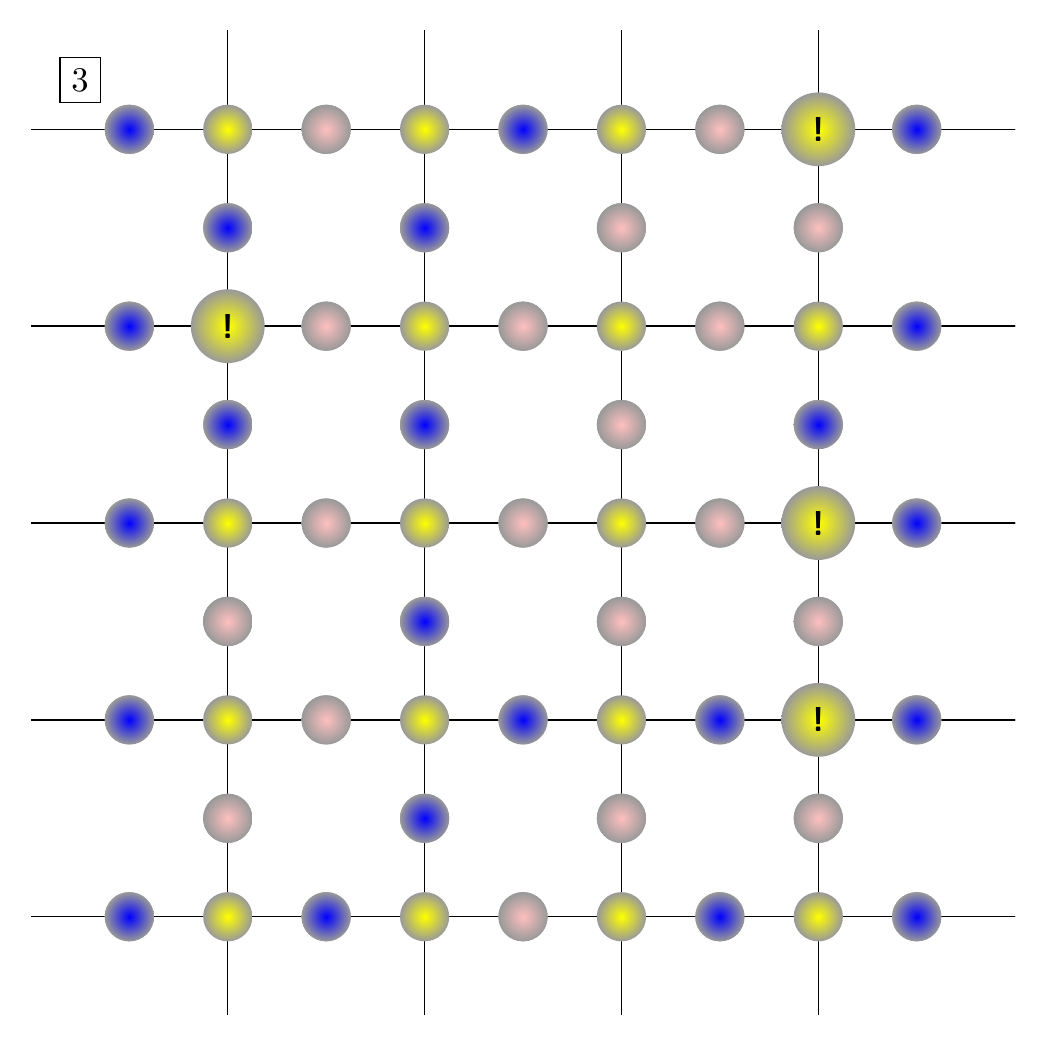}
    \vskip 0.1cm
    \includegraphics[width=0.8\columnwidth,height = 2.25in]{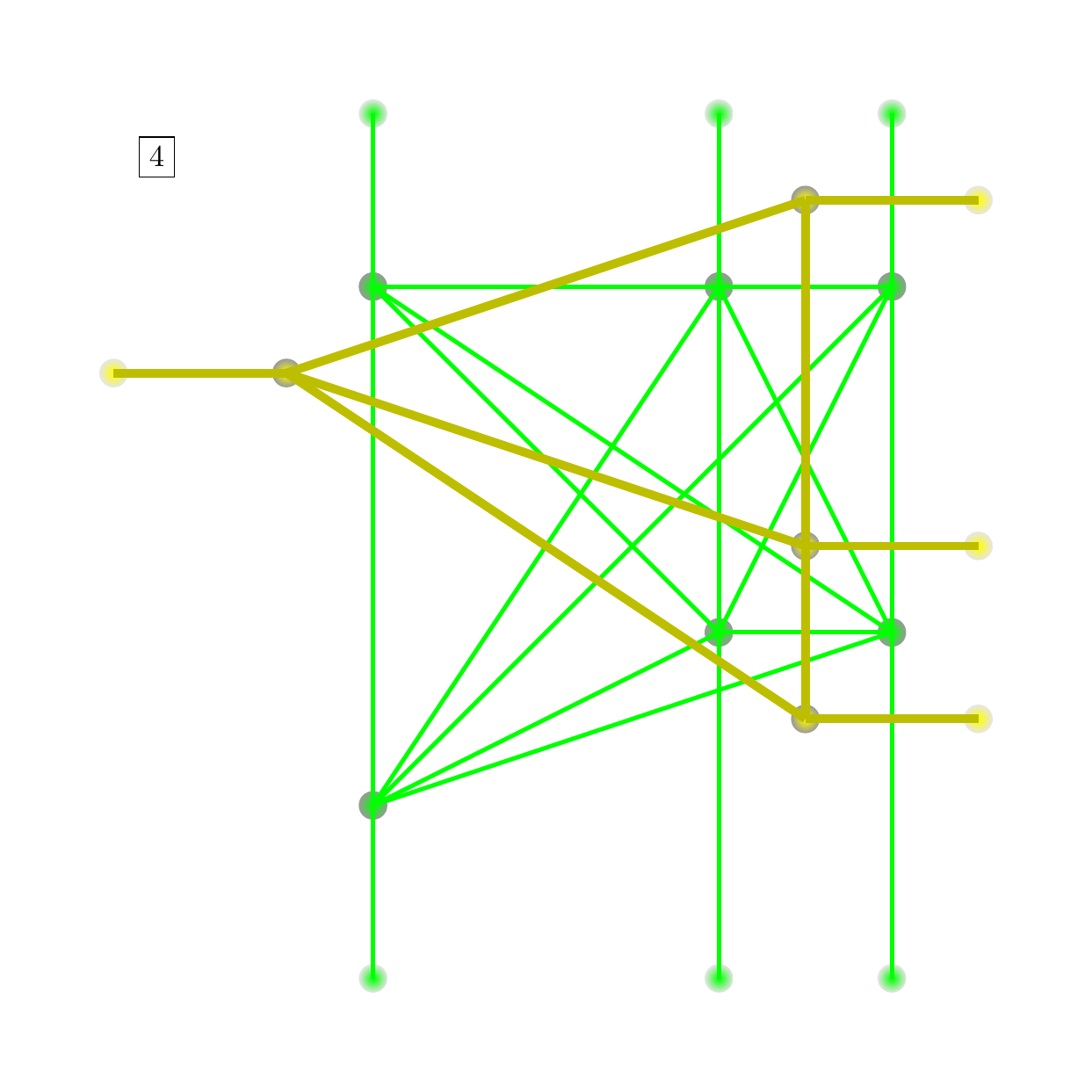}
    \vskip 0.1cm
    \includegraphics[width=0.6\columnwidth,scale=0.75]{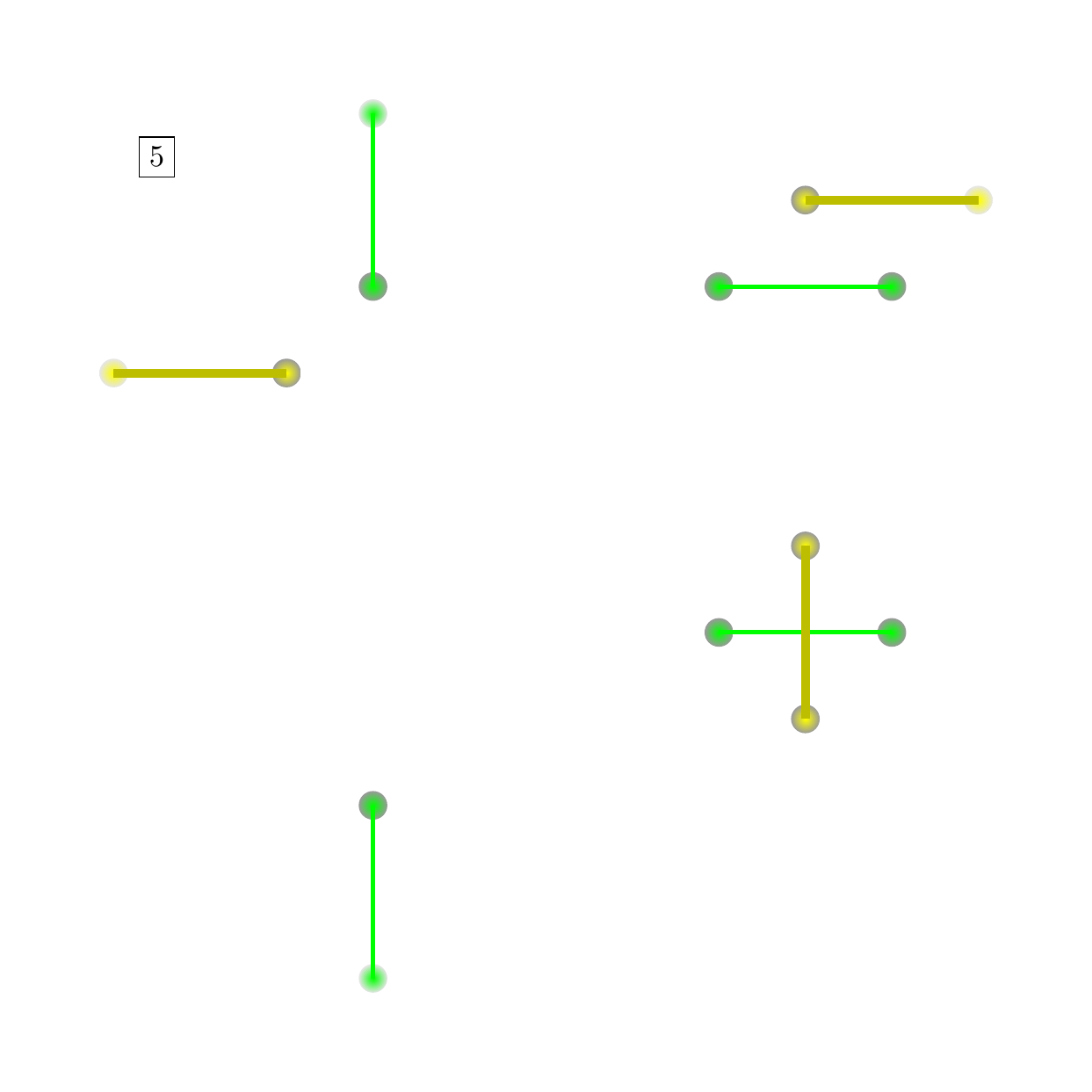}
    \includegraphics[width=0.6\columnwidth,scale=0.75]{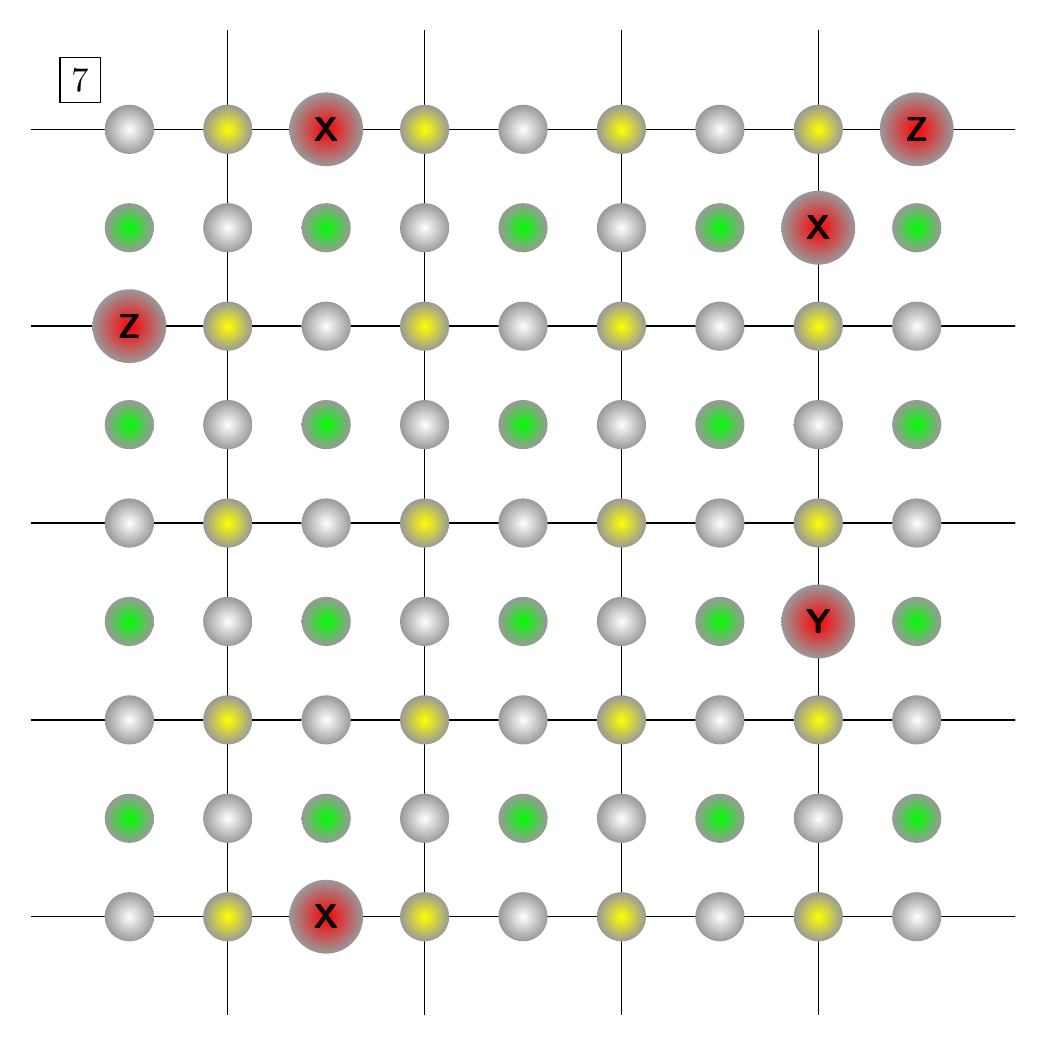}
    \vskip 0.1cm
    \includegraphics[width=0.6\columnwidth,scale=0.75]{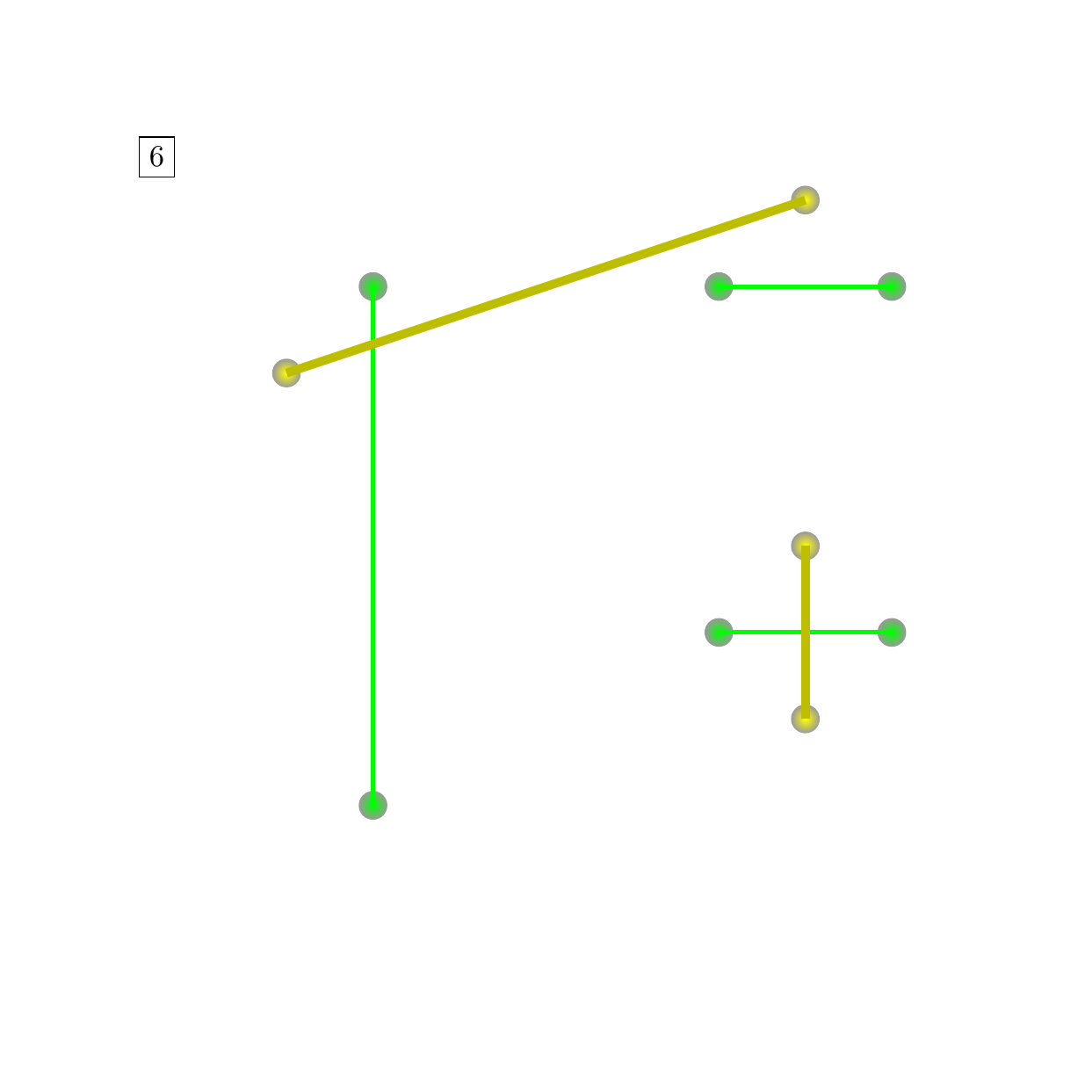}
    \includegraphics[width=0.6\columnwidth,scale=0.75]{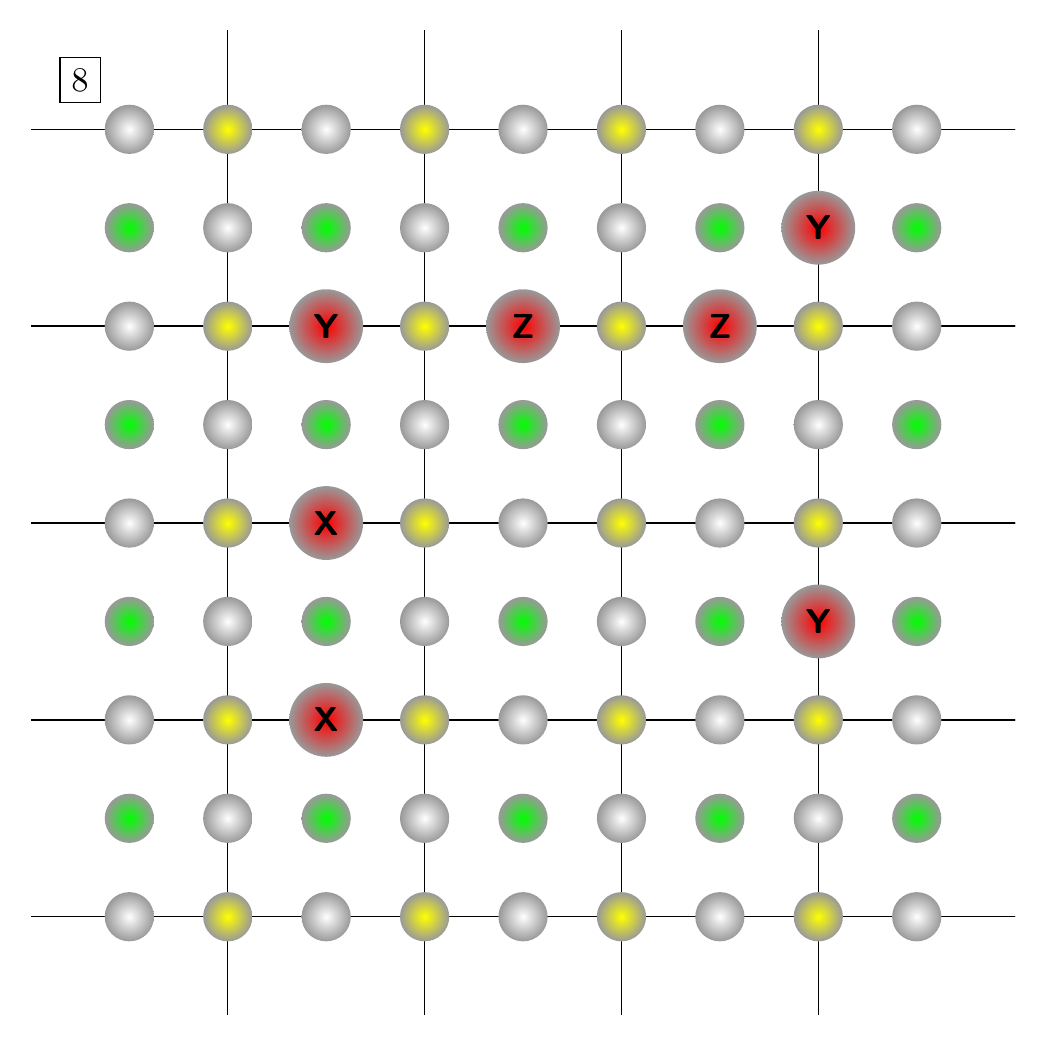}
    \caption{Example of a MWPM and rMWPM performance towards an error in a 5x5 planar code. In \textbf{1}, the proposed error composed by data qubits (white circles), data qubits experiencing non-trivial error operators (red circles), measure-z qubits and measure-x qubits (green and yellow circles). 1-syndrome measurement qubits are labeled with an exclamation mark. \textbf{2} and \textbf{3} represent the two CSS subgraphs, where pink and blue denote the qubits with lower and higher relaxation parameters respectively. \textbf{4} shows the overall graph. \textbf{5} and \textbf{6} show the minimum weight perfect matching of the graphs considering taxicab distance and reweighted distance and \textbf{7} and \textbf{8} show the recovery operators proposed by the MWPM and rMWPM decoders. }
    \label{rMWPM}
\end{figure*}

Once a surface code experiences an error, a syndrome can be extracted by measuring the measurement qubits. Since the planar code is a CSS code, the syndrome result is mapped onto two subgraphs, a Check-X subgraph, susceptible to physical $\mathrm{X}$ and  $\mathrm{Y}$ errors, and a Check-Z subgraph, susceptible to  $\mathrm{Z}$ and  $\mathrm{Y}$ errors. In both subgraphs, the respective measurement qubits act as nodes while their adjacent data qubits act as the edges that connect each measurement qubit to its four nearby measurement qubits. The MWPM decoder applied for i.i.d channels resolves the graph problem by considering all the weights of the subsequent graph to be equal, which results in ``close'' measurement qubits (syndromes) being connected via lower weight paths. However, the equal weight assumption is inappropriate when facing i.ni.d noise. Over this more restrictive channel, because each qubit suffers different levels of noise, the weights of the edges must be re-weighted according to the error parameters of the particular data qubits they represent. We refer to this modified decoding approach as re-weighted MWPM decoding. The weights we use in re-weighted MWPM (rMWPM) decoding are different for each subgraph, as each subgraph relates to a different error recovery:

\begin{equation}
    \begin{aligned}
    w_{i,X} = -\log(1-e^{-t/T_{1,i}}) \propto -\log(p_{x,i} + p_{y,i}),
    \\
    w_{i,Z} = -\log(1-e^{-t/T_{2,i}}) \propto -\log(p_{z,i} + p_{y,i}),
    \end{aligned}
\end{equation}

where $T_{1,i}$ and $T_{2,i}$ are the relaxation and dephasing parameters specific to the data qubit of the edge $i$, $t$ indicates the time that has passed since the code was initialized, and $p_i$ indicates the probability of a qubit to experience an error of type $i$. This weight consideration significantly increases the complexity of the graph problem, since the distance between two syndromes can no longer be determined through the taxicab metric \cite{degenMWPM}. Instead, we use Dijkstra's algorithm \cite{Dijstra} to determine the weight of the minimum-weight paths between syndrome 1-measurement qubits. Dijkstra's algorithm finds the shortest path between nodes in a graph and has a maximum complexity of $O(N\log(N)+M)$, where  $N$ is the number of nodes within the graph and $M$ represents the number of edges. Based on our weight convention, the weight of a chain or co-chain between two syndromes $i$ and $j$ in a subgraph $k$ will be given by:

\begin{equation}
    \sum_{k=i}^j -\log(p_{k,l}) = -\log (\prod_{k=i}^j p_{k,l}),
\end{equation}

where $p_{k,l}$ is the probability of errors susceptible to the syndromes of the $k$ subgraph for a qubit $l$. Higher failing probabilities $p_k$ will imply a lower weight and thus, longer chains and co chains with worse performing qubits will weigh less than shorter ones with better data qubits. This redistribution of weights alters the Minimum Weight Perfect Matching result and enhances the performance of the code. In FIG. \ref{rMWPM} we can see an example of how the rMWPM decoding process unfolds of the corresponding graphs. Subfigure 1 shows the error experienced by a $5\times5$ planar code along with the resulting 1-measurement qubits (indicated with exclamation marks). Subfigures 2 and 3 represent the check-X and check-Z measurement qubit subgraphs, respectively. Notice that the data qubits lay over the edges of the graph. Data qubits with high probability of failing in each subgraph are depicted as pink circles, while those qubits with longer relaxation times are depicted in blue.  Subfigure 4 portrays the overall graph comprised by the two independent check-X and check-Z subgraphs. Using different weight conventions results in different decoding outputs. This can be seen by comparing subfigures 5 and 6, which represents the result of applying a conventional MWPM decoder and that of using the rMWPM decoder, respectively. Consequently, the recovery operators proposed by each of these decoders will also different (pictures 7 and 8). The MWPM decoder has prioritized the lowest Hamming weight error while the rMWPM has accounted for the individual noise parameters of each data qubit in order to select the best-possible recovery operator.

Later on in the Results section we will see how the rMWPM decoding rule significantly improves the performance of generic MWPM decoding when facing i.ni.d. errors. Nonetheless, it must be mentioned that it does so at the expense of a higher decoding complexity. Reweighted MWPM decoding has a complexity of $O(n^3\log(n))$ while the conventional MWPM decoder has a complexity of $O(n^2\log(n))$ \cite{degenMWPM}, where $n$ represents the number of data qubits within the square planar code. The source of this increase in complexity comes from the introduction of Dijkstra's algorithm, which has $O(n\log(n))$ complexity (conventional MWPM decoding uses the taxicab distance which has complexity $O(1)$).

\section{The architecture optimization method}

In this subsection we describe the guidelines that make up another strategy that can be employed to improve the performance of planar codes when they are subjected to i.ni.d. noise. It involves re-arranging the planar code lattice qubits according to the noise they suffer. To start, consider the fact that some qubits of the lattice are less likely to experience errors than others. Naturally, this occurs because some qubits have longer relaxation and dephasing times than others. Against this backdrop, we can place the qubits on the lattice in a way in which better qubits (less likely to fail) are positioned in the most important sites, effectively minimizing the probability of harmful events (chains and co-chains).

\begin{figure}
    \centering
    \includegraphics[width=70mm,scale=1]{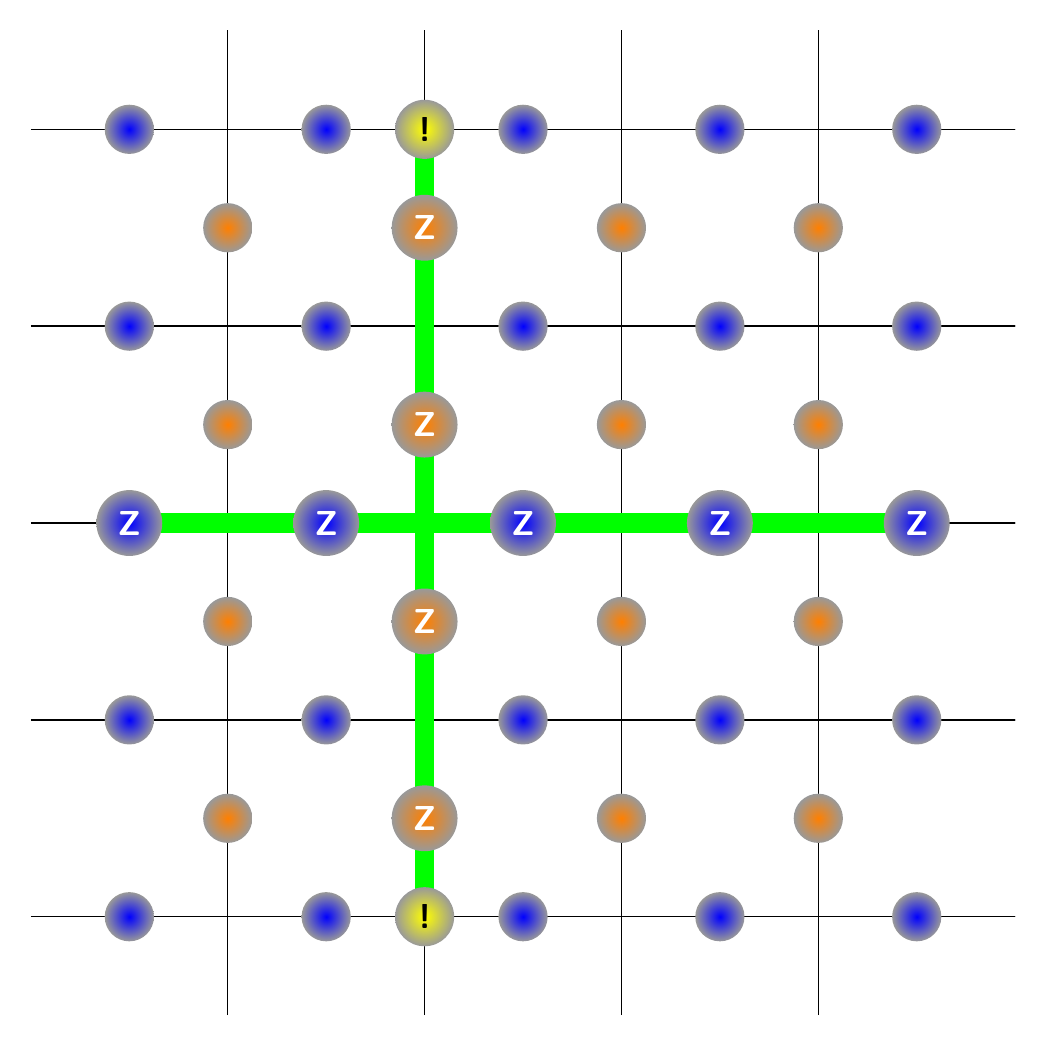}
    \caption{5x5 planar code codifying the information of a logical qubit. The data qubits of both sublattices are highlighted with identifiable blue and orange qubits. A Z-chain of weight 4 in the orange lattice is detected by two measuring qubits highlighted in yellow with an exclamation mark. Moreover, a Z chain in the blue sublattice commutes with all the measuring qubits. }
    \label{sublattices}
\end{figure}

The overall planar code lattice that encodes a logical qubit can be be split into two separate sublattices. Both of these sublattices are shown in FIG. \ref{sublattices}. The sublattice composed by qubits placed along the horizontal edges of the overall lattice does not have measurement qubits on their edges. Consequently, a horizontal $Z_i$ chain and a vertical $X_i$ co-chain will commute with all the stabilizer generators and, thus, will be in the code. On the other hand, the vertical edge sublattice has $4$ adjacent measurement qubits for all data qubits, hence, any chain or co-chain will be detected by the measurement qubits located at its endpoints.

In order to accurately re-order the qubits that make up a planar code in a way that improves performance, it is important that we come up with a way to differentiate good qubits from bad qubits. However, accurately classifying qubits according to their ``quality'' is not a simple task. Since the physical error probability that each qubit experiences varies as a function of its relaxation and dephasing times $\{T_1^j\}_{j=1}^n,\{T_2^j\}_{j=1}^n$, defining a metric that determines how good a specific qubit is with regard to the rest of the ensemble is relatively nuanced. For this reason, in our algorithm we employ the lowest relaxation time as our noise ``quality'' indicator, i.e, a larger relaxation time implies a better performing qubit. Note that dephasing is intricately related to relaxation: $1/T_2 = 1/2T_1 + 1/T_\phi$, where $T_\phi$ is the pure dephasing time \cite{TVQC}.

The minimum weight of a logical operator in an $N\times N$ planar code is $N$. For the example in FIG. \ref{sublattices}, logical operators will have a minimum weight of 5. For a logical error caused by a chain to take place, a minimum of $N$ horizontal edge qubits must experience an error. If we place the best qubits in the horizontal edge sublattice (blue in FIG. \ref{sublattices}), we will make logical error chains and co-chains more unlikely, which should ultimately lead to better code performance. Moreover, placing the worst qubits within the vertical edge sublattice (orange in FIG. \ref{sublattices}) guarantees that, at least, half of the nearest data qubits will be good ones. Based on these ideas, the algorithm we have designed to optimize the architecture of planar codes operates based on the following two principles:
\begin{itemize}
    \item  \textbf{Surround good qubits with bad qubits and vice versa}. This is done to prevent the propagation of errors through the code and the formation of chains and co-chains.
    \item \textbf{Separate differently performing qubits in both sublattices}. While the best qubits are in the $d^2$ (blue in FIG. \ref{sublattices}) sublattice, the worst ones will be in the $(d-1)^2$ (orange in FIG. \ref{sublattices}) lattice. We do this to make it unlikely for shortest weight (weight $d$) logical errors to occur. As shown in FIG. \ref{sublattices}, a chain or co-chain in the orange sublattice does not commute with the stabilizer set and, thus, placing the worst qubits within it ensures that they will not contribute towards decoding failures.
\end{itemize}

As will be shown in the next section, applying these two guidelines to re-design planar codes ends up improving their performance. In particular, we have concentrated the worst and best performing qubits in the bulk (the centre) of the code, while the most average ones have been spread out along the outer walls of the lattice. Additionally, the $(d-1)^2$ worst qubits have been placed in the orange sublattice, preventing them from contributing to the formation of minimum weight logical errors. In this manner, our method minimizes the probability of chain and co-chain formation and makes it likelier for the MWPM algorithm to be successful. Furthermore, we will also see in the Results section that the performance of the rMWPM decoder is also improved when optimizing the architecture of the code.

\section{Results}

\subsection{Planar code numerical simulation}

To estimate the performance of the various $d\times d$ planar codes with $d\in\{3,5,7\}$ \cite{surfaceReview} that we have considered in this paper we have carried out Monte Carlo numerical simulations. We have constructed the planar codes using a customized version of the QECSIM tool \cite{qecsim} that we have modified so that it can work with the i.ni.d. decoherence model. Each round of a numerical simulation is performed by generating an $N$-qubit Pauli operator, calculating its associated syndrome, and finally running the 1-cycle decoding algorithm using this syndrome as its input. Once the decoder produces an estimation of the channel error, the syndrome is extracted again. For the sake of simplicity and restrict our view to the effects of i.ni.d. noise, we will not consider measurement errors. Following this second syndrome extraction, we check that the code state commutes with the $\mathrm{X}$ and $\mathrm{Z}$ logical operators. If the second syndrome is non trivial or if the quantum state of the code no longer commutes with the logical operator, we will consider the code to have undergone a logical error. The logical error probability is obtained by computing many realizations of the aforementioned procedure. The specific properties of the constituent data qubits can be found in chapters 1 and 2 of the Supplementary material.

To estimate the logical error rater, $P_L$, of the planar codes, we choose $N_\mathrm{blocks} = 10^4$ Pauli error realizations for each considered value of the physical error probability $p$. In this way, we can guarantee that the estimated values of the logical error probability are accurate because we fulfill the Monte Carlo rule of thumb  \cite{montecarlo}
\begin{equation}
N_{\mathrm{blocks}} = \frac{100}{\mathrm{P}_L}.
\end{equation}
This rule tells us that, under the assumption that the observed error events are independent, the empirically estimated value, $\hat{\mathrm{P}}_L$, lies in the $95\%$ confidence interval of about $(0.8\hat{\mathrm{P}}_L , 1.25\hat{\mathrm{P}}_L)$. 

Finally, we estimate the average performance of the planar codes for the particular relaxation and dephasing rates of each system by performing Monte Carlo simulations in the order of $10^3$ randomized qubit arrangements defined over the particular planar code lattices.

\begin{figure*}[t!]
    \centering
    \includegraphics[width=\textwidth]{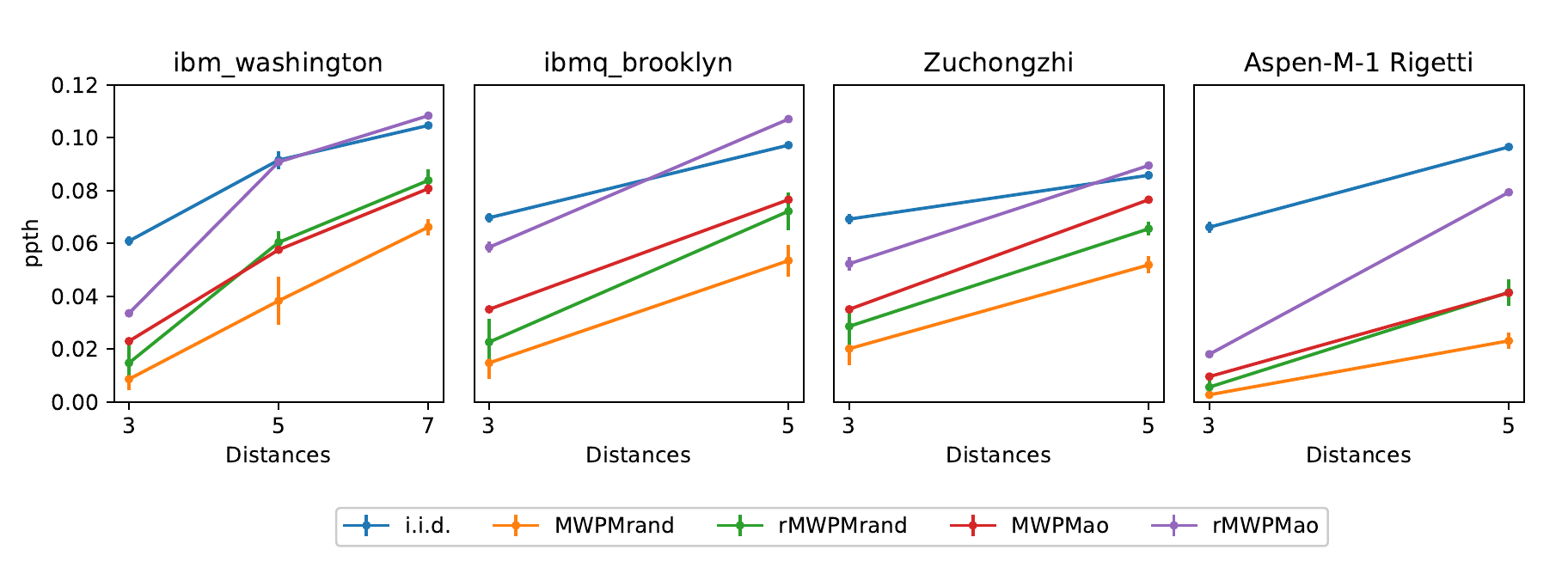}
    \caption{\textbf{Code pseudo-thresholds obtained for the qubit data of four quantum processors in different distance square planar codes.}From left to right: ibm\_washington, ibmq\_brooklyn, Zuchongzhi and Aspen-M-1 Rigetti. The blue dots represent the results under i.i.d., the yellow reflect the performance of the code under i.ni.d. using the conventional MWPM decoder. The green dots represent the performance of the code under the rMWPM decoder and the red dots represent the performance of the conventional MWPM decoder but with the code undergoing the architecture optimizing method. Lastly, the purple lines indicate the performance of the code under both architecture optimizing and the rMWPM decoder. The bars correspond to the standard deviation of all the considered configurations of the specific code.}
    \label{prm}
\end{figure*}

\subsection{Independent non identical distribution model performance}

Our simulation results are depicted in FIG. \ref{prm}. The consideration of i.ni.d. noise harshly decreases the probability pseudo threshold of all the codes; a detriment that ranges from $40\%$ to $95\%$ of the original i.i.d. noise scenario. Such a significant loss in performance is a direct consequence of assuming that drastically different qubits behave equally. This degradation may also have been exacerbated because we have considered the qubits with the highest and lowest relaxation parameters. Thus, the codes of lower distance have higher coefficients of variation. The average $T_2$ is much smaller and its coefficients of variation much higher than $T_1$, thus, the codes performance is restricted by the dephasing times of its qubits. Codes with relaxation parameters with low coefficients of variation will suffer less from i.ni.d. noise, since the individual relaxation times of the qubits will tend to be closer to the average.

\subsection{Performance of the rMWPM}

FIG. \ref{prm} shows how the rMWPM decoder outperforms the conventional MWPM decoder when i.ni.d. noise is considered. For distance 3 planar codes, its standard deviation overlaps with the MWPM stabdard deviation in all of the scenarios we have tested. This is a result of the high probability that ``bad'' qubits have of being placed in pivotal positions at such low distances, which contributes to the formation of distance-3 chains and co-chains operating non-trivially over the encoded state. Regardless, for distance 3 codes, the average performance of the rMWPM decoder exceeds that of the MWPM decoder by up to $104\%$. At distances 5 and on, the standard deviations of the rMWPM and MWPM decoder performance curves no longer overlap, but the improvement is not so significant (it ranges from $27\%$ to $79\%$). This can be understood more so as a scenario change rather than a decrease in the boost provided by the rWMPM decoder. As more qubits are used to build codes with larger distances, there will be a lower coefficient of variation between qubits (more average qubits are introduced into the lattice). Consequently, the i.ni.d. effect is not as significant as in the distance 3 scenario. 

\subsection{Performance of the architecture optimization method}

Similar to the rMWPM decoder in the previous subsection, the architecture optimization method also surpasses the performance of random data qubit layouts in all of our simulations (improvements of $22\%$ to $247\%$). In FIG. \ref{prm} we can see how at distance 3, under high coefficients of variance in $T_2$, the specific allocation of qubits within the planar code allows us to isolate the worst performing qubits. Unfortunately, because the worst qubits under perform at high rates, 1-element syndromes can end up being separated by large distances, which tricks the MWPM decoder into choosing a wrong recovery operator. As higher distances are considered and the worst qubits can be further isolated within the bulk of the surface code, the improvement in performance provided by the architecture optimization method increases (it comes close to the i.i.d. scenario in particular situations). 

When compared to the rMWPM decoder we observe that for low distance planar codes, the architecture optimization method yields better results because it successfully isolates the worst behaving qubits. As longer distance codes are tested, the rMWPM decoder gets closer and even ends up surpasses the architecture optimizer method. The exact arrangement of the qubits under the architecture optimization method for each processor can be found in the second chapter of the supplementary material section.

\subsection{Performance of the combined rMWPM and optimization method}

The true potential of the rMWPM decoder and the architecture optimization strategy comes to light when they are applied together. While the architecture optimizing method ensures that the worst qubits of the planar code are surrounded by better qubits and these bad qubits in the vertical edge bipartite lattice, the rMWPM decoder accounts for the weight asymmetry in the non trivial syndrome element graph. As a result, the combination of both of these methods produces performance increases that are $163\%$ to $650\%$ better than applying a conventional MWPM decoder over the i.ni.d. model. In some cases, the amalgamation of both of these methods surpasses the performance of the MWPM decoder under i.i.d. considerations.

\begin{figure}
    \centering
    \includegraphics[width = 0.9\columnwidth]{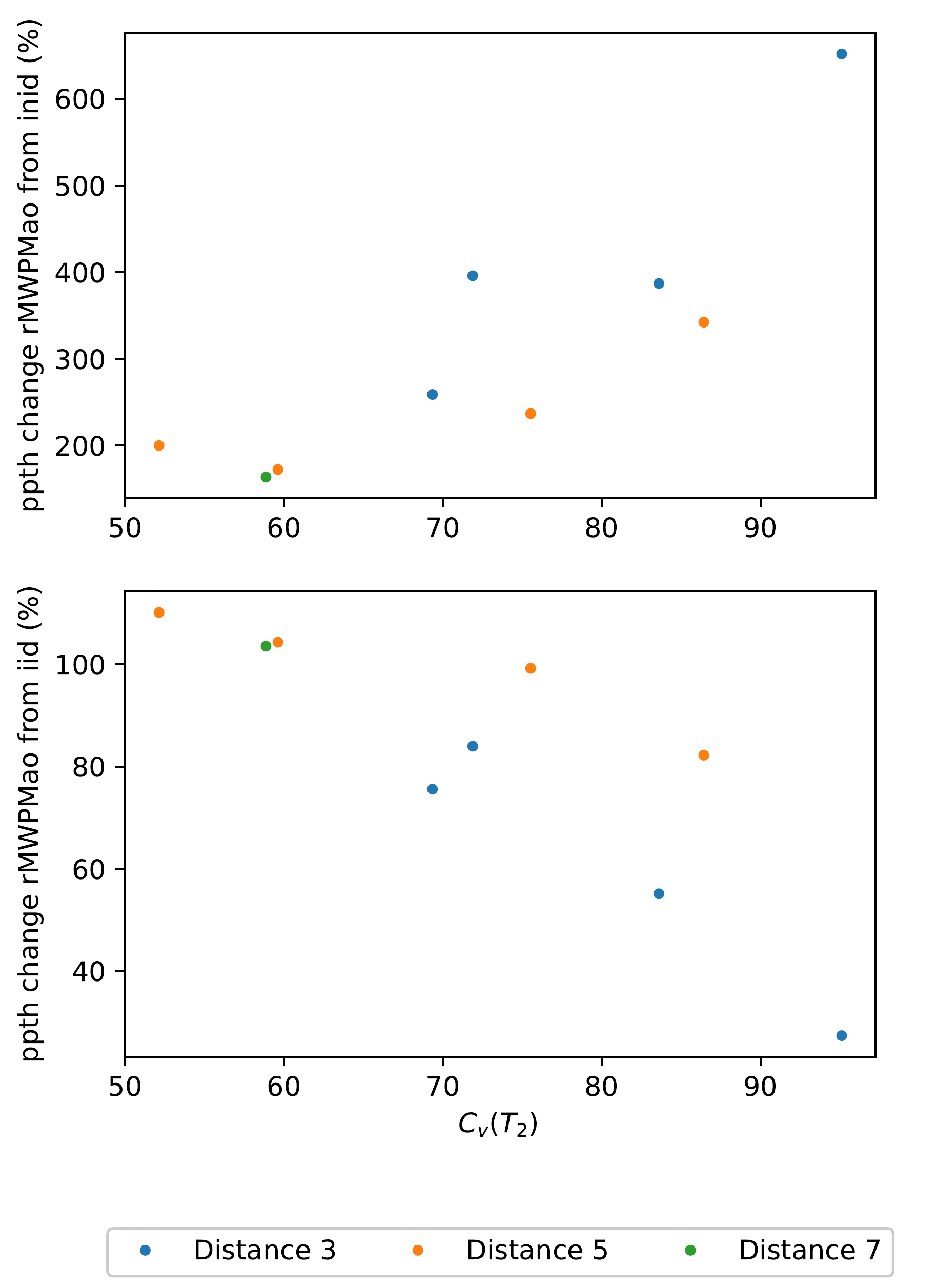}
    \caption{\textbf{Change in ppth when considering the optimized architecture and the rMWPM compared to two different scenarios with respect to the variance coefficient of $T_2$.} In top, the change in ppth is compared with the i.ni.d. case under the conventional MWPM decoder. In the bottom, it is compared with the i.i.d. case. }
    \label{ppthchange}
\end{figure}

The performance of planar codes under both conventional MWPM decoding as well as the methods we introduce in this paper is highly correlated with the coefficient of variation of the restricting relaxation parameter, $C_v(T_2)$. As can be seen in FIG. \ref{ppthchange}, a larger $C_v(T_2)$ implies a harsher drop in performance, but it also increases the yield that our methods provide. Surpassing of the i.i.d. threshold is only achieved when $C_v(T_2)$ is lower than $60\%$, where $C_v(T_2) = \sigma(T_2) / \mu(T_2)$, and $\sigma(T_2)$ and $\mu(T_2)$ are the standard deviation and average of the set of $T_2$ values.

\section{Conclusion} \label{sec:conclusion}

In this paper we have proposed the i.ni.d. noise model as an appropriate way to include the experimentally observed variance in the decoherence parameters of the qubits that make up superconducting quantum processors. Our results show how the performance of planar codes when this noise model is considered is far worse than what would be expected based on previous results obtained for the i.i.d. qubit noise model. This occurs because when the noise of the qubits of a surface code is considered to behave according to an i.ni.d noise channel, the qubits that are most likely to suffer errors (those with shorter $T_1$ and $T_2$) may form errors chains and co-chains that are too large for the decoder to successfully estimate, which ultimately results in the manifestation of logical errors.

We have also discussed how the manner in which qubits are arranged on the planar code lattice plays an important role in the performance of the codes. We saw how the ``typical'' performance of planar codes over the i.ni.d. channel is generally bad; an outcome that arises from increased likelihood that these codes have of suffering additional logical errors over the i.ni.d. paradigm (``bad'' qubits end up being located in lattice positions that cause harmful events to occur frequently). We address this issue in our work by devising two methods that tackle the inconvenience of ``low quality'' qubits. 
The first method consists in reweighting the graph on which the minimum weight perfect matching algorithm is applied. This new weight convention is directly related to the relaxation parameters and the relaxation time. In this manner, worse qubits are prioritized (over those that have larger $T_1$ and $T_2$ values) as the potential error sources.

The second technique comes down to placing qubits on the surface code lattice in a manner that guarantees that long chains and co-chains are far less probable (making use of the better qubits available). We refer to this strategy the architecture optimizing algorithm. This method enables us to prevent the placement of the worst qubits in lattice sites that are pivotal to the performance of the code. The primary working principle of the algorithm is ranking the surface code qubits according to their noise level ($T_1$ and $T_2$). Once this best-to-worst classification is defined, the ``worst'' qubits are surrounded by better qubits in order to prevent the creation of chains and co-chains, which ultimately leads to higher probabilities of successful decoding.

Both the rMWPM decoder and the architecture optmizing method improve the performance of the code significantly, an effect that is further exacerbated when the methods are combined. In fact, when they are applied jointly, performance can exceed results obtained for the i.i.d. scenario. Unfortunately, these improvements come at a price. On the one hand, rMWPM decoding can result in longer decoding times for large surface codes due to its increased complexity. For decoding to be practical, it must be performed in real time so that the noise that takes place actually  corresponds to the measured syndrome. Methods to reduce the complexity of the MWPM have been proposed \cite{degenMWPM, localMWPM}, but they also decrease its accuracy. Furthermore, experimental results show that $T_1$ and $T_2$ fluctuate both intercooldown and intracooldown \cite{TVQC,fluctGoogle,decoherenceBenchmarking,klimov,fluctAPS,fluctApp}, thus any method which is based on the knowledge of the $T$ parameters would need to be significantly faster than the life time of said changes. Additionally, the architecture optimization method has the particular drawback that, at this moment in time, it is likely that rearranging the position of the superconducting qubits that form real quantum processors once they are manufactured is not possible. It also seems unreasonable to attempt to use SWAP gates to reorder the physical qubits of the code (Note that SWAPs are constructed using CNOT gates that are really noisy at this moment in time). In any case, the results provided herein prove that the fact that each physical qubit has its own noise dynamics is critical for code performance and should not be neglected when studying quantum error correction codes.

Another takeaway from our work is the fact that in order for real planar codes (and other QEC codes for that matter) to perform at the rates promised in the literature, quantum hardware must be comprised of qubits with uniform relaxation and dephasing times. Traditionally, the literature on experimental implementation of superconducting qubits has based its elemental hardware quality claims on best-case or mean scenarios. However, it is the actual distribution of these parameters, not just the best-case or the mean values, that is most relevant to predict how good the surface codes that will operate on such hardware can be. Another possible approach is to consider that quantum systems are limited by their worst qubit and to assume that all the constituent qubits of the system behave like their ``weakest link''. However, this would be a somewhat reductionist view that would miss out on the code performance improvements that can be obtained by considering qubit differences (as is done by our architecture optimization method). Thus, by designing codes for performance over the i.ni.d. channel, one may achieve lower qubit overhead for similar code performances. This is an important outcome, since qubits are an expensive resource.

Another important issue that plays a role in the performance of real surface codes is that of gate and measurement errors in quantum hardware. We have excluded the presence of these phenomena from this work, but they are an important source of errors that should be studied. While the literature on surface codes has already considered these error sources \cite{surfaceReview}, in a similar manner to what happens for decoherence parameters, real superconducting quantum hardware will suffer different gate and measurement errors for each of their constituent qubits. This is an important problem and optimizing surface codes for such non-uniform scenarios is germane to the field of QEC.

We also believe that it is important to consider the i.ni.d. noise model for other quantum computing tasks, not just for the purpose of QEC. For example, quantum error mitigation, which is an important approach to deal with noisy quantum algorithms running on real quantum hardware, should also operate under the framework of the i.ni.d. noise model. As of today, qubit number and connectivity is not yet adequate to implement strong error correction strategies, and so quantum error mitigation techiques\footnote{Quantum error mitigation achieves error suppresion by sampling available Noisy Intermediate-Scale Quantum (NISQ) devices many times and classically post-processing these measurement outcomes.} are an important component of the modern quantum computing paradigm. It is possible that accounting for the non-uniformity of the noise levels that current qubits experience will lead to further improvements in mitigation techniques. Lastly, it may also be possible to improve the reliability of current quantum computers by compiling quantum algorithms specifically for the hardware that they will be executed on (accounting for the $T_1$ and $T_2$ values of individual qubits plus the individual gate and measurement error rates).

\section*{Data availability}
The data that supports the findings of this study is available from the corresponding authors upon reasonable request.

\section*{Code availability}
The code that supports the findings of this study is available from the corresponding authors upon reasonable request.

\section*{Author Contributions}
J.E.M. and P.F conceived the research and J.E.M proposed the model. A.dM.iO, J.E.M. and P.F. proposed the rMWPM method. A.dM.iO. proposed the architecture optimization algorithm and performed the numerical simulations. A.dM.iO., J.E.M., and P.F. analyzed the results and drew the conclusions. The manuscript was written by A.dM.iO., J.E.M., and P.F., and revised by P.M.C. and J.G.-F. The project was supervised by P.M.C. and J.G.-F.

\section*{Competing Interests}
The authors declare no competing interests.

\section*{Acknowledgements}
This work was supported by the Spanish Ministry of Economy and Competitiveness through the ADELE project (Grant No. PID2019-104958RB-C44), by the Spanish Ministry of Science and Innovation through the proyect Few-qubit quantum hardware, algorithms and codes, on photonic and solid-state systems (PLEC2021-008251) and by the Diputación Foral de Gipuzkoa through the DECALOQC Project No. E 190 / 2021 (ES). This work was funded in part by NSF Award No. CCF-2007689.

 \section*{Supplementary Material}

\section*{Supplementary Note 1: Relaxation and dephasing times of realistic quantum hardware}
In the primary text we make the claim that most of the currently-existing state-of-the-art superconducting quantum processors are made up of qubits whose relaxation and dephasing times are not the same. In what follows we present the experimental data that justifies this claim, and we discuss how these results inspired the proposal of the i.ni.d. decoherence model that we present in the main text.

\begin{figure*}[ht!]
    \centering
    \includegraphics[width=\textwidth]{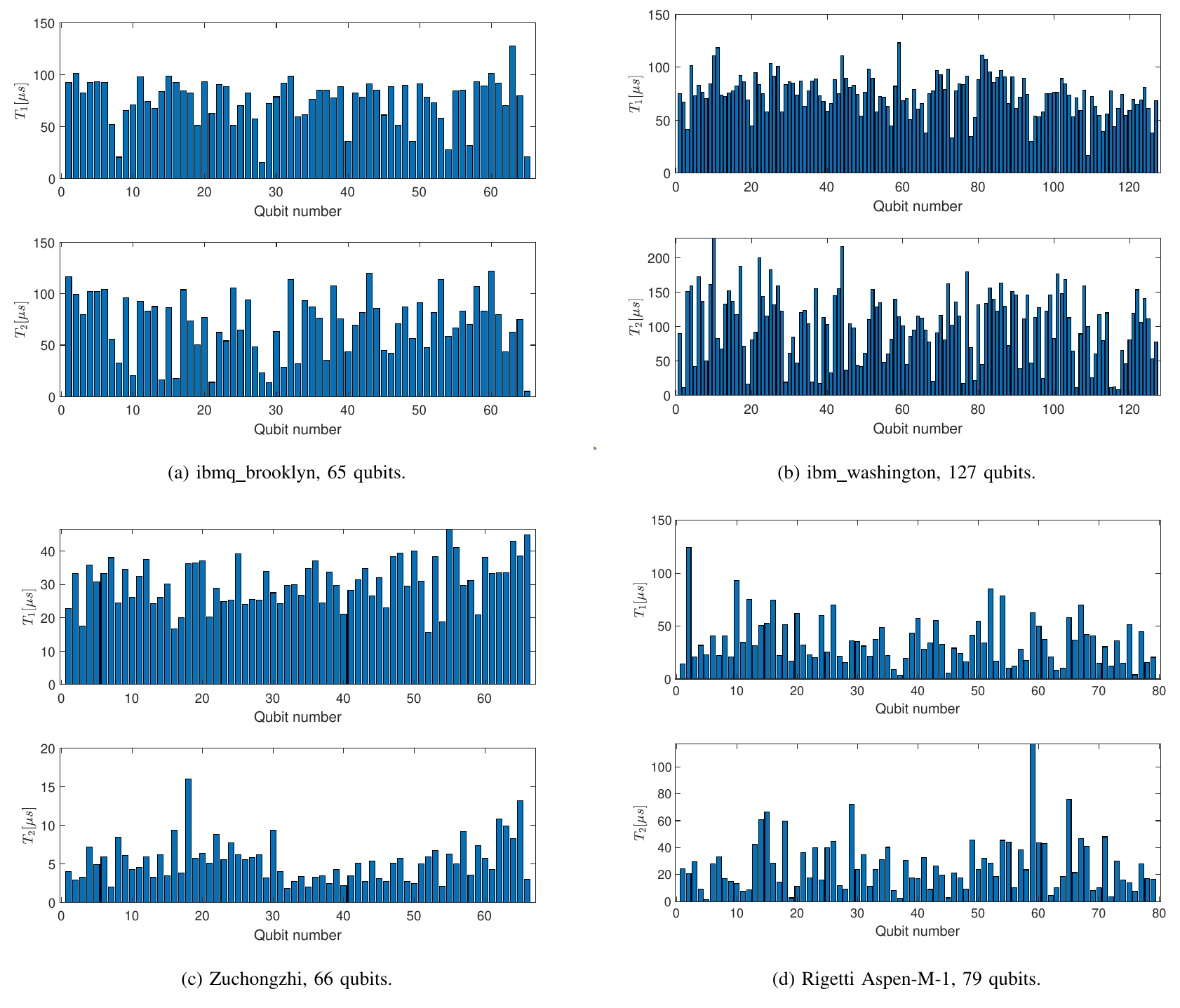}
	\caption{\textbf{Considered relaxation and dephasing times for the i.ni.d. decoherence model.} The $T_1$ and $T_2$ values associated to the individual qubits of each of the systems are presented using histograms. \textbf{a} Data for the ibmq\_brooklyn machine with 65 qubits. Timestamp: 16/11/2021 10:41 (CET) \cite{Broo}. \textbf{b} Data for the ibm\_washington machine with 127 qubits. Timestamp: 16/11/2021 12:21 (CET) \cite{Wash}. \textbf{c} Data for the Zuchongzhi machine with 66 qubits. Data from \cite{Zuchongzhi}. \textbf{d} Data for the Rigetti Aspen-M-1 machine with 38 qubits. Data obtained using the Strangeworks platform \cite{Aspen11}. Timestamp: 30/05/2022 11:37 (CET). Note that the Aspen-M-1 quantum computer is made up of 40 qubits, but when we requested the corresponding data, we were only able to obtain measurements related to of its qubits.}
	\label{experimentaldata}
\end{figure*}

Figure \ref{experimentaldata} shows the specific values of $T_1$ and $T_2$ that we have employed to simulate the performance of planar codes over the i.ni.d. channel. Recall that this channel model can account for differences in the values of $T_1$ an $T_2$ of invidividual qubits (see The independent non-identically distributed decoherence model in the main text) \cite{IBMqexp,Broo,Wash,Zuchongzhi,Aspen11}. The experimental measurements shown in Figure \ref{experimentaldata} reveal how the particular relaxation and dephasing times of each individual qubit within the quantum system can vary drastically. For example, the qubits that make up the ibm\_washington quantum processor exhibit a minimum relaxation time of $16.54$ $\mu s $ and a maximum relaxation time of $123.11$ $\mu s$, i.e, there are qubits hwose relaxation time differs by an order of magnitude. This phenomenon is further exacerbated for the ibm\_washington qubit dephasing times. The minimum dephasing time value is $8.58$ $\mu s$ and the maximum value is $228.56$ $\mu s$. This behaviour can be observed over all of the superconducting machines considered in this paper. We summarize the minimum and maximum $T_1$ and $T_2$, as well as their mean values, in Table \ref{minmaxT1T2}. The main takeaway here is that, within the real quantum systems, the decoherence parameters of each constituent qubit will vary significantly. Because this type of behaviour must be considered when building accurate decoherence models, the i.ni.d. noise model we propose in the main text is a relevant contribution to the field of QEC, as it can accurately re-enact the real quantum noise processes that experimental multi-qubit systems can suffer.

\begin{table*}[t]
\centering
\begin{tabular}{|ccccccc|}
\hline
Processor      & $\min{T_1}[\mu s]$ & $\max{T_1}[\mu s]$   & $\mu_{T_1}[\mu s]$   & $\min{T_2}[\mu s]$ & $\max{T_2}[\mu s]$   & $\mu_{T_2}[\mu s]$   \\ \hline
ibmq\_brooklyn & $15.37$ & $127.82$   & $75.3554$   & $5.06$ & $122.19$ & $70.4778$  \\
ibm\_washington & $16.54$ & $123.11$   & $74.2827$  & $8.58$ & $228.56$ & $101.4081$ \\
Zuchongzhi & $15.6$ & $46.6$ & $30.6045$   & $1.8$ & $16$ & $5.3348$  \\
Rigetti Aspen-M-1 & $3.95$ & $124.35$  & $35.77$ & $1.22$ & $117.08$ & $26.5$\\
\hline
\end{tabular}
\caption{\textbf{Minimum, maximum and mean $T_1$ and $T_2$ for the systems considered in our work.} The values are obtained from the data provided in Figure \ref{experimentaldata}.}
\label{minmaxT1T2}
\end{table*}

Some details regarding the data shown in Figure \ref{experimentaldata} merit further discussion. To start off, notice how the $T_1$ and $T_2$ values we have considered are all timestamped (refer to Figure \ref{experimentaldata}). This is related to the fact that these values can vary between calibration rounds (inter-calibration) and even during the calibration process itself (intra-calibration) \cite{fluctGoogle,decoherenceBenchmarking,klimov,fluctAPS,fluctApp}. The data that quantum computing companies make available is generally updated inter-calibration (from calibration to calibration), so it is important that we state that the data we have employed in our analysis strictly relates to different calibration cycles of various quantum machines. Because intracalibration fluctuations are not usually reported, even if they are important \cite{decoherenceBenchmarking,klimov,fluctAPS,fluctApp,TVQC}, it is not something that we have been able to consider in the present study. 

Additionally, we must also disclose that part of the available data does not make physical sense. For example, the data reported for qubit 3 of the ibm\_washington quantum processor tells us that $T_1=41.09$ $\mu s$ and that $T_2=150.47$ $\mu s$. It is physically impossible for these values to be correct, since they do not comply with the Ramsey limit $T_2\leq 2T_1$. We believe that these ``erroneous'' readings stem from the fact that $T_1$ and $T_2$ measurements are not performed during the same time instant. Because intracalibration decoherence parameter fluctuation can take place \cite{decoherenceBenchmarking,klimov,fluctAPS,fluctApp,TVQC}, it is likely that by the time the second experiment is run, the decoherence parameters have already changed. In light of this, whenever we encounter such data readings in our work, we have considered that that the qubit in question actually saturates the Ramsey limit ($T_2=2T_1$) so that our simulations can be run. This also speaks towards the importance of measuring the decoherence parameters of the qubits simultaneously, as this would produce more accurate data. Additionally, this also sheds light on the importance of understanding and characterizing intracalibration decoherence parameter fluctuation, as this phenomenon may also impact the performance of real quantum error correction codes \cite{TVQC}.

\begin{figure*}[!b]
    \centering
    \includegraphics[width=0.3\textwidth]{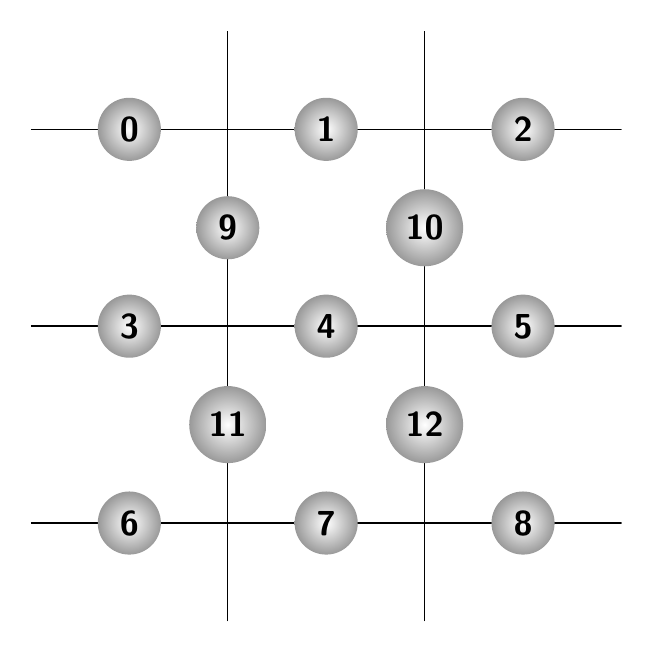}
    \includegraphics[width=0.3\textwidth]{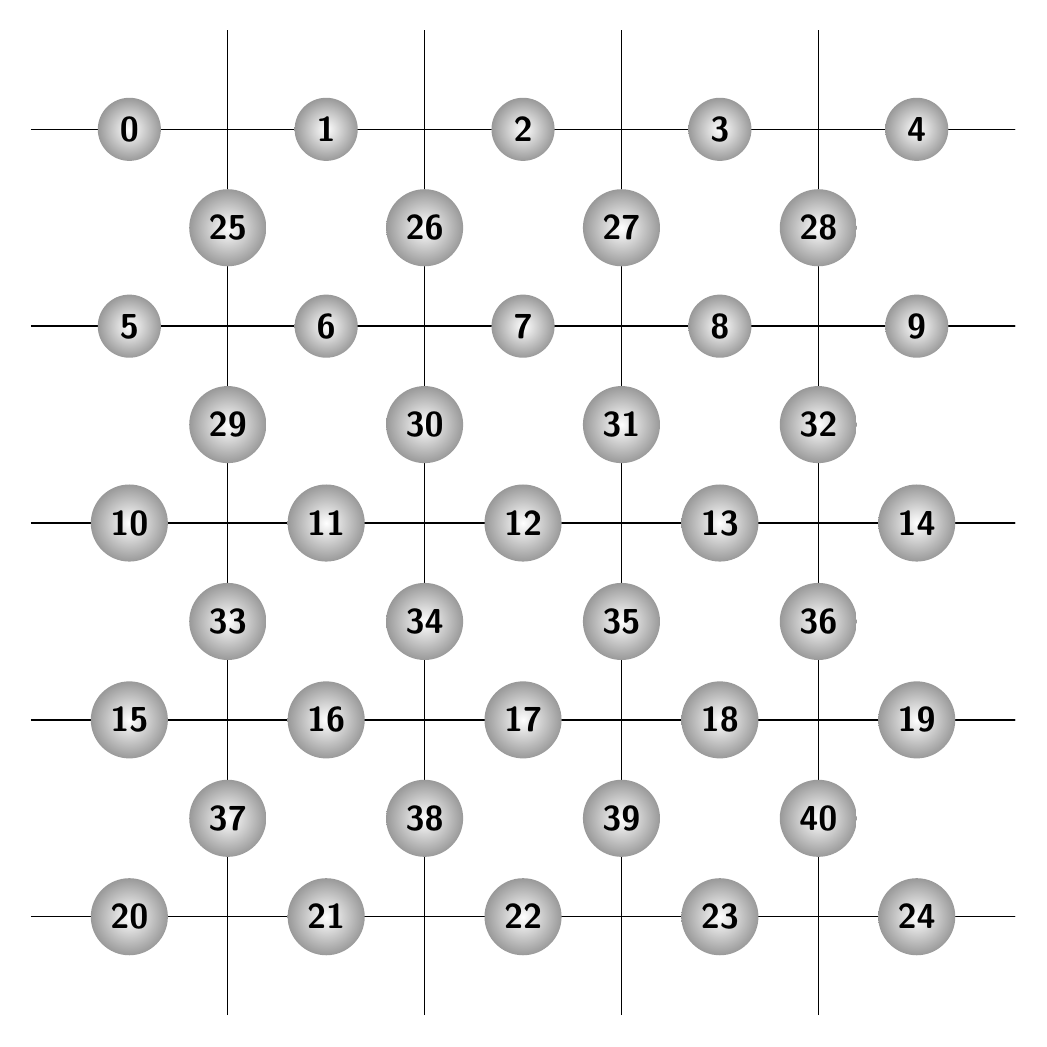}
    \includegraphics[width=0.3\textwidth]{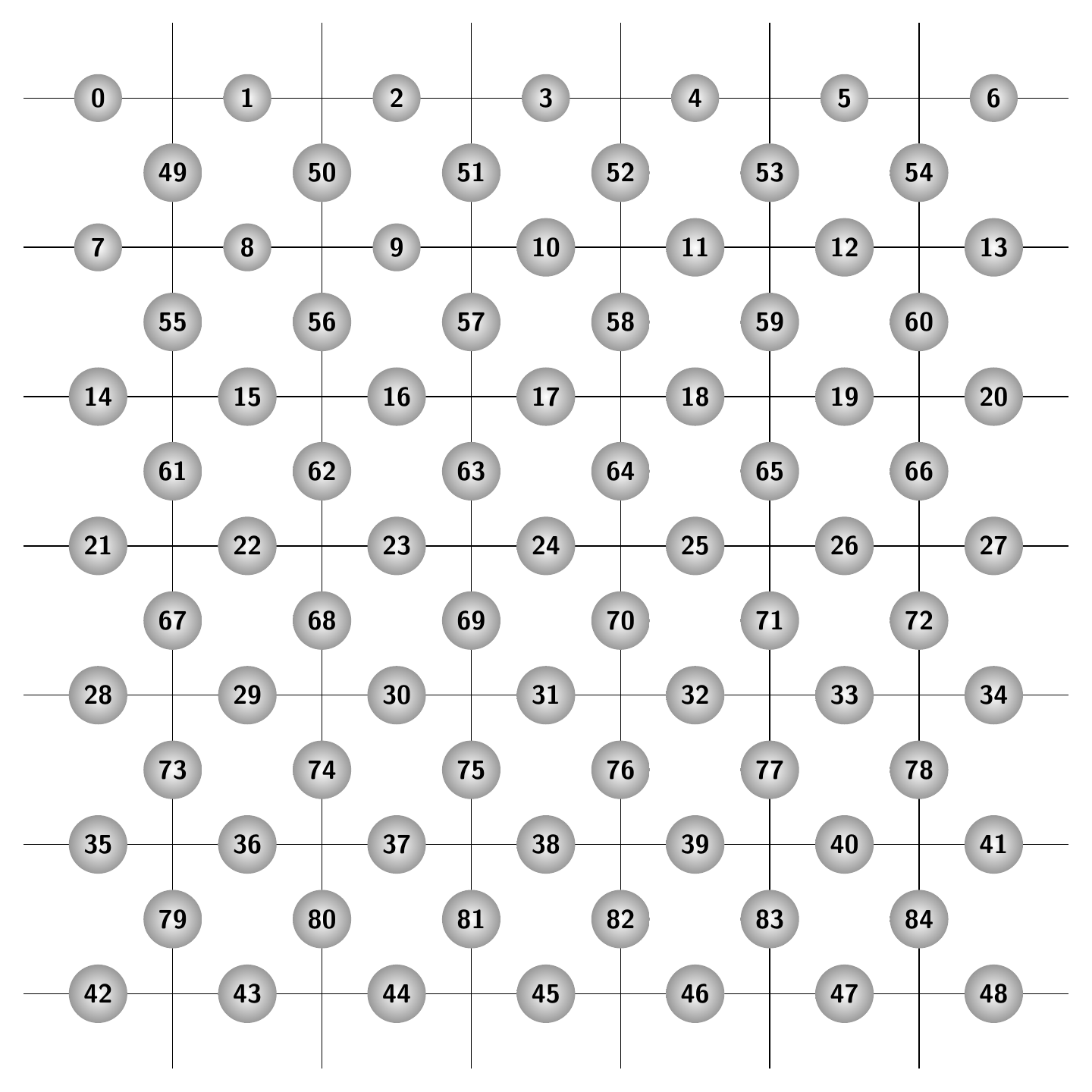}
    \caption{Graphical representation of the data qubits of 3, 5 and 7 planar codes. Each data qubit is represented by a gray-shaded dot and a number representing its index.}
    \label{indices}
\end{figure*}

\section*{Supplementary Note 2: Considered planar architectures}

In this section, we present the qubit arrangements that have been considered for the planar codes in the main text. This reveals the way in which this new architecture distributes qubits according to their $T_1$ and $T_2$ values. The location of a given qubit within the lattice is given by a combination of indices, as is done by convention in the QECSIM library \cite{qecsim}. These indices are computed based on the equation:

\begin{equation}
\begin{split}
    \text{index}_{r,c} =& (r \symbol{92} 2) * (cols - c \% 2) + \\
    &(c \symbol{92} 2) + (r \% 2 * rows * cols),    
\end{split}
\end{equation}

where \textit{cols} and \textit{rows} indicate the size of the given planar code, and $r$ and $c$ are the specific row and column of the given qubit. Additionally, the $\%$ sign denotes the modulo-$2$ operation and $\symbol{92}$ represents the integer or floor division operation. This index-based labeling system allows us to easily distinguish between qubits that are placed on the different sublattices discussed in the main text (the horizontal edge qubits are labeled first followed by the vertical edge qubits). An example of this labeling system is shown in FIG. \ref{indices}, where the qubit indices for each of the planar code lattices that have been considered in this article is shown.

Following such indexing, the results of FIG. 1\textbf{b} in the main text are based on arranging the qubits using the same indexing of the qubits of the ibm\_washington processor. The specific arrangements of the qubits obtained by the optimization algorithm are displayed in Tables \ref{3x3ibmqBrooklyn}-\ref{3x3Aspen}. The data shown in this tables corresponds to the data presented in FIG. \ref{experimentaldata} of Supplementary Note 1.

\begin{table*}[h!]
\centering
\begin{tabular}{||c |c |c||} 
 \hline
 Qubit index & $T_1$($\mu$ s) & $T_2$ ($\mu$ s)\\ [0.5ex] 
 \hline\hline
 0 & 84.1 & 16.0 \\ 
 1 & 92.9 & 104.3  \\
 2 & 93.3 & 106.9  \\
 3 & 99.0 & 113.5  \\
 4 & 101.3 & 122.2  \\ 
 5 & 101.6 & 99.1 \\ 
 6 & 93.6 & 102.4  \\
 7 & 93.0 & 101.9 \\
 8 & 92.8 & 17.6 \\
 9 & 63.0 & 13.9  \\
 10 & 21.1 & 5.1 \\
 11 & 72.2 & 13.8 \\
 12 & 15.4 & 23.3  \\ [1ex] 
 \hline
\end{tabular}
\caption{Optimized architecture for the $3\times 3$ planar code considering the ibmq\_brooklyn quantum processor.}
\label{3x3ibmqBrooklyn}
\end{table*}

\begin{table*}[h!]
\begin{tabular}{c c c}
\centering
\begin{tabular}{||c |c |c||}  
 \hline
 Qubit index & $T_1$($\mu$ s) & $T_2$ ($\mu$ s)\\ [0.5ex] 
 \hline\hline
 0 & 70.5 & 43.8 \\ 
 1 & 78.8 & 47.8  \\
 2 & 78.4 & 81.6  \\
 3 & 92.3 & 79.6  \\
 4 & 89.2 & 83.0  \\ 
 5 & 84.6 & 103.8 \\ 
 6 & 98.7 & 86.2  \\
 7 & 91.6 & 91.5 \\
 8 & 98.2 & 92.4 \\
 9 & 92.9 & 104.3  \\
 10 & 93.3 & 106.9 \\
 11 & 99.0 & 113.5 \\
 12 & 101.3 & 122.2 \\
 13 & 101.6 & 99.1 \\ [1ex] 
 \hline
\end{tabular}
\begin{tabular}{||c| c |c||}
 \hline
 Qubit index & $T_1$($\mu$ s) & $T_2$ ($\mu$ s)\\ [0.5ex] 
 \hline\hline
 14 & 93.6 & 102.4 \\
 15 & 93.0 & 101.9 \\
 16 & 92.7 & 116.1 \\
 17 & 91.6 & 119.9  \\
 18 & 89.9 & 87.2 \\ 
 19 & 85.4 & 85.5  \\
 20 & 85.1 & 83.2  \\
 21 & 82.9 & 94.0  \\
 22 & 82.4 & 79.3  \\ 
 23 & 57.8 & 48.2 \\ 
 24 & 61.3 & 44.7  \\
 25 & 35.8 & 43.3 \\
 26 & 85.0 & 35.3 \\
 27 & 31.6 & 70.3  \\ [1ex] 
 \hline
\end{tabular}
\begin{tabular}{||c |c |c||}
 \hline
 Qubit index & $T_1$($\mu$ s) & $T_2$ ($\mu$ s)\\ [0.5ex] 
 \hline\hline
 28 & 28.1 & 28.8 \\
 29 & 71.4 & 20.6 \\
 30 & 84.1 & 16.0 \\
 31 & 63.0 & 13.9 \\
 32 & 21.1 & 5.1 \\
 33 & 72.2 & 13.8 \\
 34 & 15.4 & 23.3\\
 35 & 92.8 & 17.6  \\
 36 & 20.7 & 32.8 \\ 
 37 & 92.2 & 28.5  \\
 38 & 59.6 & 32.2  \\
 39 & 35.6 & 56.6  \\
 40 & 88.5 & 42.1  \\[1ex] 
 \hline
\end{tabular}
\end{tabular}
\label{5x5ibmqBrooklyn}
\caption{Optimized architecture for the $5\times 5$ planar code considering the ibmq\_brooklyn quantum processor.}
\end{table*}

\begin{table*}[h!]
\centering
\begin{tabular}{||c |c |c||} 
 \hline
 Qubit index & $T_1$($\mu$ s) & $T_2$ ($\mu$ s)\\ [0.5ex] 
 \hline\hline
 0 & 43.6 & 12.2 \\ 
 1 & 100.7 & 159.0  \\
 2 & 103.3 & 182.8  \\
 3 & 110.9 & 228.6  \\
 4 & 123.1 & 114.0  \\ 
 5 & 111.0 & 215.8 \\ 
 6 & 107.6 & 133.7  \\
 7 & 101.8 & 158.8 \\
 8 & 16.5  & 100.0\\
 9 & 77.6 & 11.6 \\
 10 & 61.1 & 8.6  \\
 11 & 71.1 & 11.2 \\
 12 & 66.9 & 11.9 \\ [1ex] 
 \hline
\end{tabular}
\caption{Optimized architecture for the $3\times 3$ planar code considering the ibm\_washington quantum processor.}
\label{3x3ibmqWashington}
\end{table*}

 \begin{table*}[h!]
\begin{tabular}{c c c}
\centering
\begin{tabular}{||c |c |c||}  
 \hline
 Qubit index & $T_1$($\mu$ s) & $T_2$ ($\mu$ s)\\ [0.5ex] 
 \hline\hline
 0 & 65.5 & 33.0 \\ 
 1 & 34.7 & 69.4  \\
 2 & 89.8 & 111.6  \\
 3 & 96.7 & 90.7  \\
 4 & 91.1 & 130.0  \\ 
 5 & 94.9 & 91.7 \\ 
 6 & 92.0 & 187.7  \\
 7 & 95.3 & 155.8 \\
 8 & 98.3 & 110.1 \\
 9 & 100.7 & 159.0  \\
 10 & 103.3 & 182.8 \\
 11 & 110.9 & 228.6 \\
 12 & 123.1 & 114.0 \\
 13 & 111.0 & 215.8 \\ [1ex] 
 \hline
\end{tabular}
\begin{tabular}{||c| c |c||}
 \hline
 Qubit index & $T_1$($\mu$ s) & $T_2$ ($\mu$ s)\\ [0.5ex] 
 \hline\hline
 14 & 107.6 & 133.7 \\
 15 & 101.8 & 158.8 \\
 16 & 98.4 & 161.7 \\
 17 & 97.2 & 151.2  \\
 18 & 93.0 & 116.5 \\ 
 19 & 91.7 & 131.4  \\
 20 & 91.6 & 179.1  \\
 21 & 90.7 & 151.2  \\
 22 & 90.5 & 122.4  \\ 
 23 & 89.7 & 37.2 \\ 
 24 & 33.6 & 101.9  \\
 25 & 72.3 & 25.6 \\
 26 & 52.7 & 21.6 \\
 27 & 86.9 & 20.2  \\ [1ex] 
 \hline
\end{tabular}
\begin{tabular}{||c |c |c||}
 \hline
 Qubit index & $T_1$($\mu$ s) & $T_2$ ($\mu$ s)\\ [0.5ex] 
 \hline\hline
 28 & 83.5 & 17.9 \\
 29 & 68.9 & 17.0 \\
 30 & 43.6 & 12.2 \\
 31 & 77.6 & 11.6 \\
 32 & 61.1 & 8.6 \\
 33 & 71.1 & 11.2 \\
 34 & 66.9 & 11.9\\
 35 & 16.5 & 100.0  \\
 36 & 72.8 & 17.6 \\ 
 37 & 83.5 & 19.2  \\
 38 & 78.0 & 21.0  \\
 39 & 57.8 & 25.1  \\
 40 & 29.9 & 47.0  \\[1ex] 
 \hline
\end{tabular}
\end{tabular}
\caption{Optimized architecture for the $5\times 5$ planar code considering the ibm\_washington quantum processor.}
\label{5x5ibmqWashington}
 \end{table*}

\begin{table*}[h!]
\begin{tabular}{c c c}
\centering
\begin{tabular}{||c |c |c||}  
 \hline
 Qubit index & $T_1$($\mu$ s) & $T_2$ ($\mu$ s)\\ [0.5ex] 
 \hline\hline
 0 & 50.8 & 86.1 \\ 
 1 & 53.3 & 64.1  \\
 2 & 54.4 & 117.4  \\
 3 & 78.5 & 159.1  \\
 4 & 79.0 & 80.8  \\ 
 5 & 81.3 & 140.5 \\ 
 6 & 82.6 & 117.8  \\
 7 & 83.1 & 98.4 \\
 8 & 83.8 & 199.4 \\
 9 & 84.4 & 161.0  \\
 10 & 84.8 & 84.5 \\
 11 & 86.8 & 120.0 \\
 12 & 88.5 & 144.8 \\
 13 & 89.3 & 147.7 \\
 14 & 89.8 & 111.6 \\
 15 & 96.7 & 90.7 \\
 16 & 91.1 & 130.0 \\
 17 & 94.9 & 91.7  \\
 18 & 92.0 & 187.7 \\ 
 19 & 95.3 & 155.8  \\
 20 & 98.3 & 110.1  \\
 21 & 100.7 & 159.0  \\
 22 & 103.3 & 182.8  \\ 
 23 & 110.9 & 228.6 \\ 
 24 & 123.1 & 114.0  \\
 25 & 111.0 & 215.8 \\
 26 & 107.6 & 133.7 \\
 27 & 101.8 & 158.8  \\
 28 & 98.4 & 161.7 \\
 29 & 97.2 & 163.5 \\ [1ex] 
 \hline
\end{tabular}
\begin{tabular}{||c| c |c||}
 \hline
 Qubit index & $T_1$($\mu$ s) & $T_2$ ($\mu$ s)\\ [0.5ex] 
 \hline\hline
 30 & 93.0 & 116.5 \\ 
 31 & 91.7 & 131.4  \\
 32 & 91.6 & 179.1  \\
 33 & 90.7 & 151.2  \\
 34 & 90.5 & 122.4  \\ 
 35 & 89.6 & 154.0 \\ 
 36 & 88.7 & 155.3  \\
 37 & 88.4 & 131.1 \\
 38 & 85.8 & 140.0 \\
 39 & 84.4 & 115.5  \\
 40 & 84.1 & 168.6 \\
 41 & 118.4 & 83.2 \\
 42 & 82.9 & 172.6 \\
 43 & 82.0 & 140.0 \\
 44 & 81.3 & 104.0 \\
 45 & 78.9 & 95.2 \\
 46 & 55.6 & 119.9 \\
 47 & 53.7 & 113.4  \\
 48 & 52.9 & 127.3 \\ 
 49 & 71.7 & 48.4  \\
 50 & 54.5 & 46.1  \\
 51 & 70.4 & 44.7  \\
 52 & 74.4 & 44.5  \\ 
 53 & 54.1 & 42.0 \\ 
 54 & 41.1 & 150.5  \\
 55 & 71.8 & 38.8 \\
 56 & 38.1 & 95.0 \\
 57 & 34.7& 39.4  \\
 58 & 65.5 & 33.0 \\
 59 & 72.3 & 25.6 \\ [1ex] 
 \hline
\end{tabular}
\begin{tabular}{||c |c |c||}
 \hline
 Qubit index & $T_1$($\mu$ s) & $T_2$ ($\mu$ s)\\ [0.5ex] 
 \hline\hline
 60 & 52.7 & 21.6 \\ 
 61 & 86.9 & 20.2  \\
 62 & 83.5  & 17.9  \\
 63 & 68.9 & 17.0  \\
 64 & 43.6 & 12.2  \\ 
 65 & 77.6 & 11.6 \\ 
 66 & 61.1 & 8.6  \\
 67 & 71.1 & 11.2 \\
 68 & 66.9 & 11.9 \\
 69 & 16.5 & 100.0  \\
 70 & 72.8 & 17.6 \\
 71 & 83.5 & 19.2 \\
 72 & 78.0 & 21.0 \\
 73 & 57.8 & 25.1 \\
 74 & 29.9 & 47.0 \\
 75 & 33.6 & 101.9 \\
 76 & 89.7 & 37.2 \\
 77 & 38.1 & 53.0  \\
 78 & 39.6 & 80.0 \\ 
 79 & 73.3 & 41.9  \\
 80 & 44.3 & 80.4  \\
 81 & 44.7 & 81.7  \\
 82 & 111.7 & 44.8  \\ 
 83 & 74.0 & 47.4 \\ 
 84 & 70.1 & 50.7  \\ [1ex] 
 \hline
\end{tabular}
\end{tabular}
\caption{Optimized architecture for the $7\times 7$ planar code considering the ibm\_washington quantum processor.}
\label{7x7ibmqWashington}
\end{table*}

 \begin{table*}[h!]
\centering
\begin{tabular}{||c |c |c||} 
 \hline
 Qubit index & $T_1$($\mu$ s) & $T_2$ ($\mu$ s)\\ [0.5ex] 
 \hline\hline
 0 & 21.1 & 2.2 \\ 
 1 & 29.8 & 9.2  \\
 2 & 27.5 & 9.4  \\
 3 & 33.4 & 10.8  \\
 4 & 36.2 & 16.0\\ 
 5 & 38.5 & 13.2 \\ 
 6 & 33.5 & 9.9  \\
 7 & 16.7 & 9.4 \\
 8 &  33.7 & 2.5 \\
 9 & 34.8 & 2.0\\
 10 & 29.7 & 1.8 \\
 11 & 38.0 & 2.0 \\
 12 & 18.8 & 2.1 \\ [1ex] 
 \hline
\end{tabular}
\caption{Optimized architecture for the $3\times 3$ planar code considering the Zuchongzhi quantum processor.}
\label{3x3Zuchongzhi}
\end{table*}

 \begin{table*}[h!]
 \begin{tabular}{c c c}
\centering
\begin{tabular}{||c |c |c||}  
 \hline
 Qubit index & $T_1$($\mu$ s) & $T_2$ ($\mu$ s)\\ [0.5ex] 
 \hline\hline
 0 & 24.2 & 3.3 \\ 
 1 & 26.8 & 3.4\\
 2 & 15.6 & 5.9 \\
 3 & 26.1 & 6.2  \\
 4 & 25.3 & 6.2 \\ 
 5 & 37.1 & 6.4 \\ 
 6 & 35.8 & 7.2  \\
 7 & 25.2 & 7.7\\
 8 & 24.4 & 8.5 \\
 9 & 29.8 & 9.2  \\
 10 & 27.5 & 9.4 \\
 11 & 33.4 & 10.8 \\
 12 & 36.2 & 16.0 \\
 13 & 38.5 & 13.2 \\ [1ex] 
 \hline
\end{tabular}
\begin{tabular}{||c| c |c||}
 \hline
 Qubit index & $T_1$($\mu$ s) & $T_2$ ($\mu$ s)\\ [0.5ex] 
 \hline\hline
 14 & 33.5 & 9.9 \\
 15 & 16.7 & 9.4 \\
 16 & 28.8 & 8.8 \\
 17 & 42.9 & 8.3  \\
 18 & 20.9 & 7.4 \\ 
 19 & 38.3 & 6.7  \\
 20 & 46.6 & 6.3  \\
 21 & 39.2 & 6.2 \\
 22 & 34.5 & 6.1  \\ 
 23 & 30.1 & 3.5 \\ 
 24 & 37.0 & 3.3  \\
 25 & 34.0 & 3.2 \\
 26 & 44.9 & 3.0 \\
 27 & 29.4 & 2.7  \\ [1ex] 
 \hline
\end{tabular}
\begin{tabular}{||c |c |c||}
 \hline
 Qubit index & $T_1$($\mu$ s) & $T_2$ ($\mu$ s)\\ [0.5ex] 
 \hline\hline
 28 & 34.7 & 2.7 \\
 29 & 39.9 & 2.5\\
 30 & 21.1 & 2.2 \\
 31 & 34.8 & 2.0 \\
 32 & 29.7 & 1.8 \\
 33 & 38.0 & 2.0\\
 34 & 18.8 & 2.1\\
 35 & 33.7 & 2.5 \\
 36 & 30.0 & 2.7 \\ 
 37 & 22.9 & 2.7  \\
 38 & 33.3 & 2.9  \\
 39 & 32.1 & 3.1  \\
 40 & 17.5 & 3.3  \\[1ex] 
 \hline
\end{tabular}
\label{5x5Zuchongzhi}
\end{tabular}
\caption{Optimized architecture for the $5\times 5$ planar code considering the Zuchongzhi quantum processor.}
 \end{table*}

\begin{table*}[h!]
\centering
\begin{tabular}{||c |c |c||} 
 \hline
 Qubit index & $T_1$($\mu$ s) & $T_2$ ($\mu$ s)\\ [0.5ex] 
 \hline\hline
 0 & 12.5 & 3.2 \\ 
 1 & 78.8 & 45.4  \\
 2 & 50.7 & 60.6  \\
 3 & 52.9 & 66.3  \\
 4 & 62.3 & 117.1\\ 
 5 & 58.2 & 76.1 \\ 
 6 & 51.2 & 60.0 \\
 7 & 69.6 & 47.0 \\
 8 & 4.0  & 7.6\\
 9 & 16.9 & 2.7 \\
 10 & 23.2 & 1.2 \\
 11 & 3.9 & 2.5  \\
 12 & 5.37 & 2.7 \\ [1ex] 
 \hline
\end{tabular}
\caption{Optimized architecture for the $3\times 3$ planar code considering the Rigetti Aspen-M-1 quantum processor.}
\label{3x3Aspen}
\end{table*}

  \begin{table*}[h!]
 \begin{tabular}{c c c}
\centering
\begin{tabular}{||c |c |c||}  
 \hline
 Qubit index & $T_1$($\mu$ s) & $T_2$ ($\mu$ s)\\ [0.5ex] 
 \hline\hline
 0 & 10.0 & 20.1 \\ 
 1 & 12.1 & 10.2\\
 2 & 85.1 & 28.7 \\
 3 & 30.5 & 48.0  \\
 4 & 48.6 & 31.2 \\ 
 5 & 34.0 & 32.1 \\ 
 6 & 36.1 & 72.1  \\
 7 & 42.35 & 40.1\\
 8 & 50.2 & 43.6 \\
 9 & 78.9 & 45.4  \\
 10 & 50.8 & 60.6 \\
 11 & 52.9 & 66.3 \\
 12 & 62.3 & 117.1 \\
 13 & 58.24 & 76.1 \\ [1ex] 
 \hline
\end{tabular}
\begin{tabular}{||c| c |c||}
 \hline
 Qubit index & $T_1$($\mu$ s) & $T_2$ ($\mu$ s)\\ [0.5ex] 
 \hline\hline
 14 & 21.2 & 59.9 \\
 15 & 69.7 & 46.9 \\
 16 & 69.7 & 44.5 \\
 17 & 41.4 & 45.8  \\
 18 & 37.6 & 42.9 \\ 
 19 & 32.3 & 36.1  \\
 20 & 31.7 & 42.6  \\
 21 & 31.2 & 34.9 \\
 22 & 36.0 & 30.2  \\ 
 23 & 10.3 & 18.6 \\ 
 24 & 15.1 & 10.1  \\
 25 & 16.3 & 9.2 \\
 26 & 34.4 & 8.9 \\
 27 & 9.0 & 8.2  \\ [1ex] 
 \hline
\end{tabular}
\begin{tabular}{||c |c |c||}
 \hline
 Qubit index & $T_1$($\mu$ s) & $T_2$ ($\mu$ s)\\ [0.5ex] 
 \hline\hline
 28 & 8.2 & 10.4 \\
 29 & 20.8 & 4.6\\
 30 & 12.6 & 3.2 \\
 31 & 17.0 & 2.7 \\
 32 & 23.2 & 1.2 \\
 33 & 4.0 & 2.6\\
 34 & 5.4 & 2.7\\
 35 & 4.0 & 7.6 \\
 36 & 34.6 & 7.5 \\ 
 37 & 41.0 & 8.2  \\
 38 & 75.5 & 8.8  \\
 39 & 31.9 & 9.0  \\
 40 & 15.8 & 9.2  \\[1ex] 
 \hline
\end{tabular}
\label{Rigetti}
\end{tabular}
\caption{Optimized architecture for the $5\times 5$ planar code considering the Rigetti Aspen-M-1 quantum processor.}
 \end{table*}
 
 \clearpage


\begin{thebibliography}{00}



\bibitem{qudits}
Wang, Y., Hu, Z., Sanders, B. C. \& Kais, S. Qudits and High-Dimensional Quantum Computing. \emph{Front. Phys.} \textbf{8,} (2020).

\bibitem{qudits2}
Low, P. J., White, B. M., Cox, A. A., Day, M. L. \& Senko, C. Practical trapped-ion protocols for universal qudit-based quantum computing. \emph{Phys. Rev. Research} \textbf{2,} 033128 (2020).

\bibitem{CVQC}
Mile, G., Weedbrook, C., Menicucci, N. C., Ralph, T. C. \& van Loock, P. Quantum computing with continuous-variable clusters. \emph{Phys. Rev. A} \textbf{79,} 062318 (2009).

\bibitem{CVQC2}
Bourassa, J. E. et al. Blueprint for a Scalable Photonic Fault-Tolerant Quantum Computer. \emph{Quantum} \textbf{5,} 392 (2021).

\bibitem{NielsenChuang} Nielsen, M. A. \& Chuang, I. Quantum Computation and Quantum Information: 10th Anniversary Edition, (Cambridge Univ. Press, Cambridge, 2011).

\bibitem{josurev} Etxezarreta Martinez, J., Fuentes, P., Crespo, P. M. \& Garcia-Fr\'ias, J. Approximating Decoherence Processes for the Design and Simulation of Quantum Error Correction Codes on Classical Computers. \emph{IEEE Access} \textbf{8}, 172623-172643 (2020).

\bibitem{TVQC} Etxezarreta Martinez, J., Fuentes, P., Crespo, P. M., \& Garcia-Fr\'ias, J. Time-varying Quantum Channel Models for Superconducting Qubits. \emph{npj Quantum Inf.} \textbf{7,} 115 (2021).

\bibitem{gatetimes} Tomesh, T. et al. SupermarQ: A Scalable Quantum Benchmark Suite. \emph{arXiv preprint} (2022).

\bibitem{QSC} Gottesman, D. Stabilizer codes and quantum error correction. \emph{Ph.D. dissertation}, California Inst. Tech., Pasadena, CA, USA (1997).

\bibitem{bicycle} MacKay, D. J. C., Mitchinson, G. \& McFadden P. L. Sparse-graph
codes for quantum error correction. \emph{IEEE Trans. Inf. Theory} \textbf{50}, 2315-2330 (2004).

\bibitem{qldpc15} Babar, Z., Botsinis, P., Alanis, D., Ng, S. X. \& Hanzo, L. Fifteen Years of Quantum LDPC Coding and Improved Decoding Strategies. \emph{IEEE Access} \textbf{3,} 2492-2519 (2015).

\bibitem{patrick} Fuentes, P., Etxezarreta Martinez, J., Crespo, P. M. \& Garcia-Fr\'ias, J. An approach for the construction of non-CSS LDGM-based quantum codes. \emph{Phys. Rev. A} \textbf{102}, 012423 (2020).

\bibitem{patrick2} Fuentes, P., Etxezarreta Martinez, J., Crespo, P. M., \& Garcia-Fr\'ias, J. Design of LDGM-based quantum codes for asymmetric quantum channels. \emph{Phys. Rev. A} \textbf{103,} 022617 (2021).

\bibitem{QTC} Poulin, D., Tillich, J.-P. \& Ollivier, H. Quantum Serial Turbo Codes. \emph{IEEE Trans. Inf. Theory} \textbf{55}, 2776-2798 (2009).

\bibitem{EAQTC} Wilde, M. M., Hsieh, M. \& Babar, Z. Entanglement-Assisted Quantum Turbo Codes. \emph{IEEE Trans. Inf. Theory} \textbf{60}, 1203-1222 (2014).

\bibitem{josu} Etxezarreta Martinez, J., Crespo, P. M. \& Garcia-Fr\'ias, J. Depolarizing Channel Mismatch and Estimation Protocols for Quantum Turbo Codes. \emph{Entropy} \textbf{21(12)}, 1133 (2019).

\bibitem{josu2} Etxezarreta Martinez, J., Fuentes P., Crespo,P. M. \& Garcia-Fr\'ias, J. Pauli Channel Online Estimation Protocol for Quantum Turbo Codes. \emph{2020 IEEE International Conference on Quantum Computing and Engineering (QCE)}, 102-108 (2020).

\bibitem{qecsim}
Tuckett, D. K. Tailoring surface codes: Improvements in quantum error correction with biased noise. \emph{Ph.D. Thesis}, University of Sydney, Australia, https://qecsim.github.io (2020).

\bibitem{toric} Kitaev, A. Y. Quantum computations: Algorithms and error correction. \emph{Russ. Math. Surveys} \textbf{52,} 6 (1997).

\bibitem{IBMqexp}
IBM Quantum. https://quantum-computing.ibm.com/services, 2022.

\bibitem{Wash}
ibmWashington IBM Quantum. https://quantum-computing.ibm.com/services, 2022.

\bibitem{Broo}
ibmqBrooklyn IBM Quantum. https://quantum-computing.ibm.com/services, 2022.

\bibitem{Zuchongzhi}
Wu, Y. et al. Strong Quantum Computational Advantage Using a Superconducting Quantum Processor. \emph{Phys. Rev. Lett.} \textbf{127,} 180501 (2021).

\bibitem{Aspen11}
Rigetti Aspen-M-1. Data obtained through Strangeworks. https://app.quantumcomputing.com, 2022.

\bibitem{surfaceReview} Fowler, A. G.,  Mariantoni, M., Martinis, J. M. \& Cleland, A. N. Surface codes: Towards practical large-scale quantum computation. \emph{Phys. Rev. A} \textbf{86,} 032324 (2012).

\bibitem{twirlcorrectable} Silva, M., Magesan, E., Kribs, D. W. \& Emerson, J. Scalable protocol for identification of correctable codes. \emph{Phys. Rev. A} \textbf{78,} 1 (2008).

\bibitem{pseudoth}
Azad, U., Lipińska, A., Mahato, S., Sachdeva, R., Bhoumik, D. \& Majumdar, R. Surface Code Design for Asymmetric Error Channels. online: arXiv:2111.01486 (2021).

\bibitem{pseudoth2}
Yoder, T. J. \& Kim, I. H. The surface code with a twist. \emph{Quantum} \textbf{1,} 2 (2017).

\bibitem{fluctGoogle} Carroll, M., Rosenblatt, S., Jurcevic, P., Lauer, I. \& Kandala, A. Dynamics of superconducting qubit relaxation times. \emph{arXiv preprint} (2021).

\bibitem{decoherenceBenchmarking} Burnett, J. J. et al. Decoherence benchmarking of superconducting qubits. \emph{npj Quantum Inf.} \textbf{5,} 54 (2019).

\bibitem{klimov} Klimov, P. V. et al. Fluctuations of Energy-Relaxation Times in Superconducting Qubits \emph{Phys. Rev. Lett.} \textbf{121,} 090502 (2018).

\bibitem{fluctAPS}  Schl\"or, S. et al. Correlating Decoherence in Transmon Qubits: Low Frequency Noise by Single Fluctuators \emph{Phys. Rev. Lett.} \textbf{123,} 190502 (2019).

\bibitem{fluctApp} Stehli, A. et al. Coherent superconducting qubits from a subtractive junction fabrication process. \emph{Appl. Phys. Lett.} \textbf{117,} 124005 (2020).

\bibitem{degenMWPM}
Higgott, O. PyMatching: A Python package for decoding quantum codes with minimum-weight perfect matching. \emph{arXiv preprint} (2021).

\bibitem{MWPM}
Edmonds, J. Paths, Trees, and Flowers. \emph{Canadian Journal of mathematics} \textbf{17,} 449-467 (1965).

\bibitem{MWPMFowler}
Fowler, A. G. Minimum weight perfect matching of fault-tolerant topological quantum error correction in average $O(1)$ parallel time. \emph{Quantum Inf. Comput.} \textbf{15,} 145-158 (2015).

\bibitem{degen} Fuentes, P., Etxezarreta Martinez J., Crespo, P. M., \& Garcia-Fr\'ias, J. Degeneracy and its impact on the decoding of sparse quantum codes. \emph{IEEE Access} \textbf{9,} 89093-89119 (2021).

\bibitem{Dijstra} Dijkstra, E. W., A note on two problems in connexion with graphs. \emph{Numerische mathematik}, \textbf{1,} 269--271 (1959)

\bibitem{montecarlo} Jeruchim, M. Techniques for Estimating the Bit Error Rate in the Simulation of Digital Communication Systems. \emph{IEEE J. Sel. Areas Commun.} \textbf{2,} 153–170 (1984).

\bibitem{localMWPM} Higgot O. and Breuckmann N. P. Subsystem Codes with High Thresholds by Gauge Fixing and Reduced Qubit Overhead. \emph{Phys. Rev. X} \textbf{11,} (2021)



\end{thebibliography}
\end{document}